\documentclass[trackchanges,twocolumn]{aastex7}

\shortauthors{Mondal et al.}

\begin{document}

\newcommand{\qpah}{$q_{\rm PAH}$ }

\title{PAH emission from star-forming and quiescent galaxies at $z=0.5-0.8$ probed by JWST/MIRI}

\author[orcid=0000-0003-4531-0945, gname=Chayan, sname='Mondal']{Chayan Mondal}
\affiliation{S. N. Bose National Centre for Basic Sciences,
Block-JD, Sector-III, Salt Lake, Kolkata-700106, India}
\affiliation{Academia Sinica Institute of Astronomy and Astrophysics (ASIAA), No. 1, Section 4, Roosevelt Road, Taipei 106319, Taiwan}
\email[show]{cmondal@bose.res.in}

\author[orcid=0000-0002-3805-0789, gname=Chian-Chou, sname='Chen']{Chian-Chou Chen}
\affiliation{Academia Sinica Institute of Astronomy and Astrophysics (ASIAA), No. 1, Section 4, Roosevelt Road, Taipei 106319, Taiwan}
\affiliation{East Asian Observatory, 660 N. A’ohoku Pl., Hilo, HI 96720, USA}
\email[show]{ccchen@asiaa.sinica.edu.tw}

\author[0000-0002-8768-9298]{Kanak Saha}
\affiliation{Inter-University Centre for Astronomy and Astrophysics, Ganeshkhind, Post Bag 4, Pune 411007, India}
\email{kanak@iucaa.in}

\author[0000-0001-8490-6632]{Thomas S.-Y. Lai}
\affiliation{IPAC, California Institute of Technology, 1200 East California Boulevard, Pasadena, CA 91125, USA}
\email{ThomasLai.astro@gmail.com}

\author[0000-0003-4702-7561]{Irene Shivaei}
\affiliation{Centro de Astrobiología (CAB), CSIC-INTA, Carretera de Ajalvir km 4, Torrejón de Ardoz 28850, Madrid, Spain}
\email{ishivaei@cab.inta-csic.es}

\author[0000-0002-4205-9567]{Hiddo S. B. Algera}
\affiliation{Academia Sinica Institute of Astronomy and Astrophysics (ASIAA), No. 1, Section 4, Roosevelt Road, Taipei 106319, Taiwan}
\email{hsbalgera@asiaa.sinica.edu.tw}

\author[0000-0003-2213-7983]{Bovornpratch Vijarnwannaluk}
\affiliation{Academia Sinica Institute of Astronomy and Astrophysics (ASIAA), No. 1, Section 4, Roosevelt Road, Taipei 106319, Taiwan}
\email{bkvijarnwannaluk@asiaa.sinica.edu.tw}

\begin{abstract}
We investigate emission from polycyclic aromatic hydrocarbon (PAH) dust molecules in 224 MIRI-selected star-forming and quiescent galaxies at $z=0.5-0.8$ in the GOODS-South field using multi-wavelength data from the AstroSat/UVIT, JWST, HST, Herschel, and MUSE. We define an observed diagnostic parameter, f$_{\rm PAH}$, as the ratio of JWST/MIRI F1280W to the average of F770W and F2550W band fluxes, and demonstrate that it effectively traces the PAH mass fraction (q$_{\rm PAH}$) by capturing 7.7$\mu$m PAH emission at our targeted redshift. We observe a bimodal distribution of similar nature in both f$_{\rm PAH}$ and q$_{\rm PAH}$, suggesting that PAH enrichment in galaxies goes through a rapid transition phase from a PAH-poor to a PAH-rich system, at about 70\% solar metallicity. For the star-forming galaxies (SFGs; i.e., $\sim$78\% of the total sample), we find that f$_{\rm PAH}$ increases with stellar mass and rest-frame FUV luminosity, identifying massive, UV-bright galaxies to have higher PAH mass fraction. However, the majority of the massive quiescent galaxies (QGs) show lower PAH mass fraction relative to that of SFGs of similar mass, suggesting PAH and smaller dust grains are more susceptible to destruction during quenching processes. Among SFGs, we observe an anti-correlation between f$_{\rm PAH}$ and the O32 emission line ratio, suggesting PAH photo-dissociation in strongly ionized ISM. This anti-correlation, along with a correlation between f$_{\rm PAH}$ and gas-phase metallicity 12+log(O/H), indicates that low-mass, metal-poor SFGs, which are increasingly common at even higher redshifts, host more extreme conditions less favorable for PAH production and survival.  
\end{abstract}

\keywords{\uat{Galaxies}{573} --- \uat{Interstellar medium}{847} --- \uat{Polycyclic aromatic hydrocarbons}{1280}}

\section{Introduction} 
\label{sec:introduction}

Interstellar dust is one of the fundamental baryonic components of galaxies. Although it has a negligible contribution to the total mass of the interstellar medium (ISM), the presence of interstellar dust can significantly alter the observable light from galaxies \citep{silva1998,draine2011}. A large fraction of ultraviolet (UV) and optical photons, emitted by stars, can be absorbed or scattered away following their interaction with dust grains present in the ISM \citep{mathis1977,cardelli1989,gordon1997,calzetti2000,salim2020}. The interaction also heats up the dust particles, which further re-radiate their energy in the far-infrared wavelength as a form of thermal emission. The equilibrium temperature attained by dust depends on both the grain size distribution and the energy of the incident photons, which together decide the shape of the thermal continuum emission from dust \citep{draine2011}. In addition to this continuum emission, distinct spectral emission features arise in the mid-infrared (MIR) wavelength range due to the presence of small carbonaceous dust molecules known as the polycyclic aromatic hydrocarbons (PAHs; \citealt{tielens2008,Li2020}). These molecules typically consist of $\sim$ 10 to a few hundred carbon atoms arranged in aromatic ring structures and can contribute up to $\sim$ 15\% of the total carbon content in galaxies \citep{alan1989, tielens2008}. PAHs are primarily excited by high-energy UV photons through a single-photon stochastic heating process, leading to vibrational transitions that produce prominent emission bands centered at $\lambda \sim$ 3.3, 6.2, 7.7, 8.6, 11.3, 12.7, and 17.1 $\mu$m in the mid-infrared wavelength \citep{Li2020}. The emission from PAHs can contribute up to around 20\% of the total dust luminosity in galaxies, making it an important component for astrophysical studies \citep{smith2007}.

As PAH emission is closely linked with the presence of young, massive stars, which supply the required UV photons for their excitation, the PAH luminosity serves as a powerful tracer of the star formation rate of galaxies \citep{peeters2004,shipley2016,xie2019,lai2020,zhangho2023,ronayne2024,kim2024}. Besides, PAHs also play a crucial role in regulating the thermal balance and ionization state of the ISM through their interaction with starlight \citep{bakes1994,helou2001,berne2022}. PAH molecules also act as an important catalyst in the formation of molecular hydrogen (H$_2$) in the galaxy ISM \citep{page2009,boschman2015}. Given their multifaceted importance, studying the formation, excitation, and destruction mechanisms of PAH molecules becomes essential to understanding galaxy evolution across cosmic time.

In the Spitzer era, observation of PAH emission in galaxies advanced significantly (see \citealt{Li2020} for a review). The Multiband Imaging Photometer for Spitzer (MIPS; \citealt{rieke2004}) instrument has enabled large-scale MIR surveys from the local universe to $z\sim3$ \citep{kennicutt2003,lefloch2005,perez2005,reddy2008,dale2009,munoz2015}, providing a rich dataset which allowed PAH investigations in diverse galaxy populations \citep{elbaz2005,engelbracht2005}. Targeted observations with Spitzer's Infrared Spectrograph (IRS; \citealt{houck2004}) have reported PAH detection even up to redshift $\sim$ 4.0 \citep{lutz2005,yan2005,riechers2014}. However, the relatively coarse angular resolution of Spitzer limited its ability to study a larger population of MIR-faint distant galaxies. The advent of the Mid-Infrared Instrument (MIRI; \citealt{rieke2015}) housed in the JWST \citep{gardner2023} has provided a transformative step forward to study the MIR sky \citep{lai2022,armus2023,alberts2024,lyu2025,backhaus2025,alberts2026}. The unprecedented angular resolution, sensitivity, and availability of multiple MIR photometric bands and spectrographs of MIRI have enabled the study of PAH emission in both local and distant galaxies with improved completeness and spatial accuracy \citep{garcia2022,egorov2023,chastenet2023,sandstrom2023,shivaei2024,kim2024,garcia2024,ronayne2024}. Parallel attempts are also made to incorporate/upgrade PAH recipes in different theoretical dust emission models \citep{draine2007,dale2014,rigopoulou2021,draine2021,hensley2023,narayanan2023}, which are crucial to interpret the observed feature of PAH emission bands.

Using both photometric and spectroscopic observations, several studies have explored how the PAH mass fraction evolves with different global galaxy properties. The total stellar mass and star formation rate of galaxies are reported to show a positive correlation with PAH mass fraction and PAH luminosity \citep{shim2023,shivaei2024,chastenet2025}. PAH content in both local and distant galaxies shows an increasing trend with increasing amount of molecular gas \citep{cortzen2019,shivaei2024b}. One of the most prominent correlations of PAH mass fraction is reported with gas-phase metallicity. PAH abundance in both distant galaxies as well as across different regions within nearby galaxies increases with metallicity \citep{engelbracht2005,wu2006,aniano2020,shim2023,chastenet2023,shivaei2024}, suggesting carbon-based molecules are closely tied to the chemical enrichment of galaxies. Some studies have also reported a threshold metallicity value of 12 + log(O/H) $\sim$ 8.2 - 8.5, a regime more commonly associated with low-mass high-redshift galaxies, below which \qpah starts declining significantly while maintaining a nearly constant value above the threshold \citep{whitcomb2024,egorov2025,shivaei2024}. \citet{whitcomb2024} further noticed that variation in metallicity could influence the PAH grain size distribution. They found a relatively higher abundance of smaller PAH dust grains towards progressively metal-poor ISM, which they attribute to inhibited growth of dust grains at low metallicity. The decline of \qpah below the threshold metallicity could be connected to reduced PAH formation in metal-poor (i.e., carbon-poor) environments \citep{draine2007} or ineffective shielding of dust molecules from harsh radiation fields in metal-poor ISM \citep{madden2006,xie2019,egorov2025}. The survival of PAH molecules is also strongly linked with the ambient ISM ionizing conditions. By measuring the strength of ISM ionization through different optical and near-infrared line ratios across star-forming regions of nearby galaxies, several studies reported that smaller PAH molecules are photo-dissociated when exposed to a harder radiation field \citep{allain1996,montillaud2013,chastenet2019,lai2023,egorov2023,egorov2025,lai2025}. \citet{shivaei2017} studied 476 distant galaxies from the cosmic noon and reported PAH scarcity in galaxies with increasingly higher ionization parameter and lower metallicity.

A comprehensive investigation of PAH emission across diverse galaxy populations and cosmic epochs is essential to understand the processes that govern the formation and destruction of PAH molecules. As noted above, several recent studies have utilized JWST/MIRI photometric observations to examine PAH emission in both nearby galaxies and those at higher redshifts (predominantly at $z>1.0$). Some studies have also acquired MIR spectra using the JWST MIRI/MRS instrument to directly probe PAH bands in relatively nearby galaxies \citep{garcia2022,garcia2024,rigopou2024,zhang2024}, including a handful of objects between redshift 0.5 and 1.0 \citep{young2023,mckinney2026}.

In this work, we focus on 224 MIR-selected galaxies, which include both star-forming and quiescent populations, within the redshift range 0.5 -- 0.8, a regime that remains relatively unexplored in PAH studies. This redshift interval, which corresponds to around 2 Gyr of cosmic time, connects the cosmic noon to the present day, making it critical for constructing a comprehensive picture of PAH evolution during the period when the cosmic star formation rate density was in the declining phase \citep{zavala2021}. We employ multi-wavelength photometry spanning 38 bands from the near-ultraviolet to the far-infrared (Table \ref{table_bands}), including all eight MIRI bands from the Systematic Mid-infrared Instrument Legacy Extragalactic Survey (SMILES; \citealt{alberts2024}), enabling improved constraints on PAH emission in galaxies from a 34 square arcmin patch in the GOODS-South \citep{giavalisco2004} region (Figure \ref{fig:ds9}). We utilize the deep near-UV observations of the AstroSat Deep Field South (AUDFs; \citealt{saha2024}) from the Ultra-Violet Imaging Telescope (UVIT; \citealt{kumar2012}) onboard AstroSat to probe the rest-frame FUV emission, essential for exciting PAH molecules, of our sample galaxies down to an observed luminosity of M$_{\rm UV}$ $\sim-$16.0 mag (also reported by \citealt{bhattacharya2025}), which is comparable to the depth reached by the Hubble Deep UV Legacy Survey (HDUV; \citealt{oesch2018}). Our study demonstrates a robust method to probe 7.7$\mu$m PAH emission in galaxies within the targeted redshift range and to analyze the PAH mass fraction of selected star-forming and quiescent galaxies in relation to their physical properties, such as stellar mass, FUV luminosity, metallicity, ISM ionization, specific SFR, and observed UV continuum slope. 

The paper is arranged as follows: we discuss data and sample selection in Section \ref{sec:data_selection}, analysis in Section \ref{sec:analysis}, which includes the construction of a diagnostic parameter and CIGALE SED modeling of sample galaxies, and results and discussion in Section \ref{sec:results_and_discussion}, followed by summary in Section \ref{sec:summary}. We adopt a cosmology with $H_0 = 70$ km~s$^{-1}$ Mpc$^{-1}$, $\Omega_{\Lambda} = 0.7$, $\Omega_M = 0.3$\ throughout the paper.

\section{Data and Sample selection} 
\label{sec:data_selection}
\begin{figure}
    \includegraphics[width=3.5in]{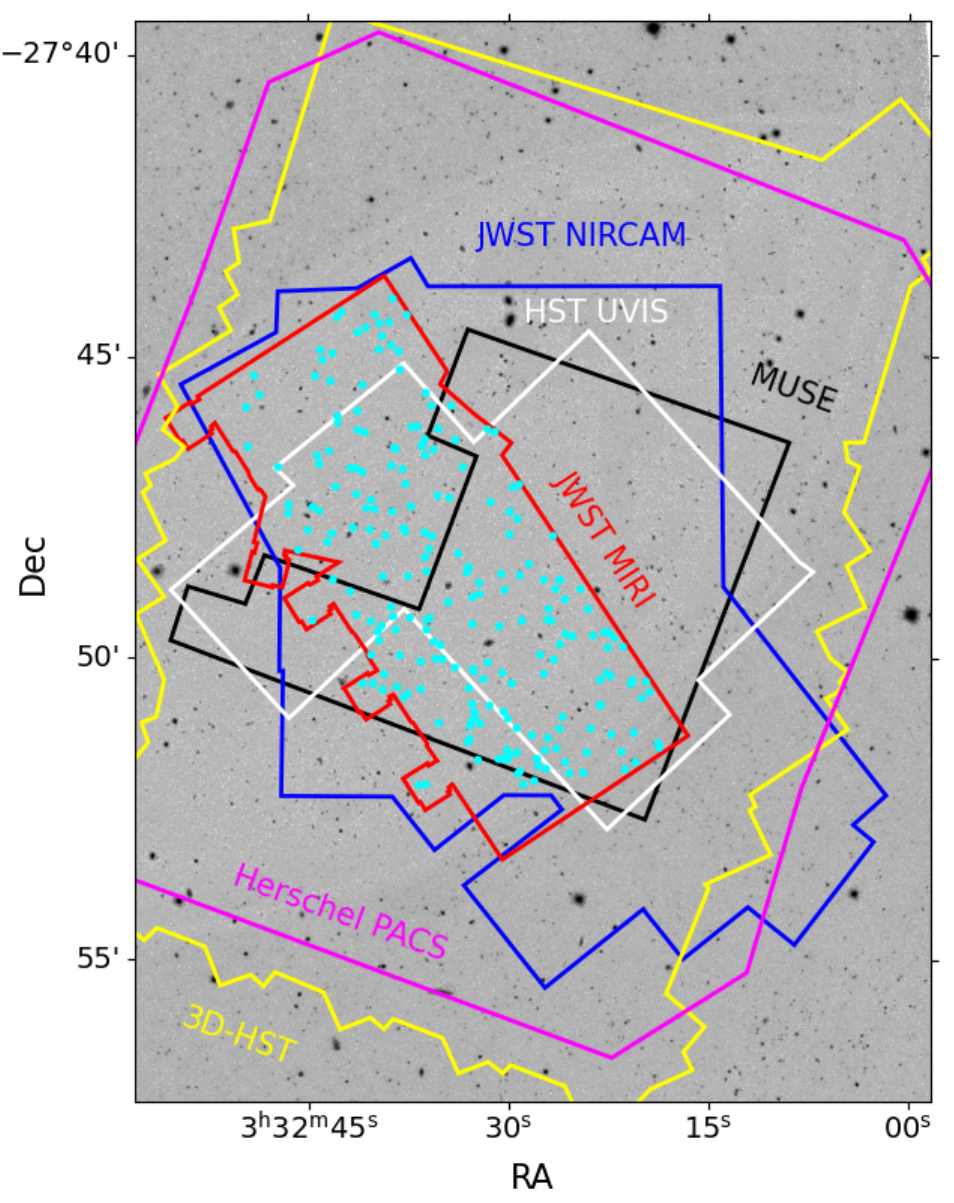}
    \caption{The location of 224 sample galaxies is shown using cyan points on the NUV image of the AstroSat UV Deep Field south (i.e., GOODS-South region) acquired from \citet{saha2024}. The sky coverages of JWST NIRCam (JADES \citealt{rieke2023}), MIRI (SMILES \citealt{alberts2024}, HST UVIS (HDUV \citealt{oesch2018}, 3DHST \citep{momcheva2016}, Herschel PACS \citep{elbaz2005}, and MUSE \citep{urrutia2019}, from which the data have been utilized in this study, are shown using polygons of different colors as indicated in the figure.}
    \label{fig:ds9}
\end{figure}

In this study, we target galaxies within the redshift range of 0.5 -- 0.8 in the GOODS-South field to understand the dependency of PAH mass fraction on different galaxy properties. As one of our primary objectives is to explore PAH emission in relation to the stellar UV radiation, sampling the rest-frame FUV and MIR emission of galaxies becomes our key requirement. We use the photometric catalog from the JWST SMILES \citep{alberts2024,rieke2024} to probe the rest-frame MIR emission that covers important PAH emission bands (eg, 3.3, 6.2, 7.7, and 11.3 $\mu$m) at our targeted redshift window. SMILES is one of the largest programs conducted with eight bands of JWST's MIRI instrument \citep{rieke2015} to cover $\sim$34 square arcmin area in the GOODS-South/Hubble Ultra Deep Field, reaching a typical 5$\sigma$ depth of 25.6 mag (F560W) to 20.8 mag (F2550W) \citep{alberts2024}. To probe the rest-frame FUV emission at this redshift, we utilize the AstroSat UV Deep Field South (AUDFs; \citealt{saha2024}) source catalogs that provide measurements in both FUV and NUV bands in the GOODS-South region. The observation of the AUDFs is conducted using F154W (FUV) and N242W (NUV) bands of the Ultra-Violet Imaging Telescope \citep{kumar2012}, reaching a 3$\sigma$ depth of $\sim$27.19 and 27.73 mag, respectively, in the deeper part (Figure \ref{fig:ds9}; \citealt{saha2024}). At redshift $\sim0.65$, UVIT's N242W filter, used in this study, covers the rest-frame FUV emission (i.e., wavelength $\sim$1200 -- 1800 $\AA{}$). For the rest-frame optical and infrared photometry, we utilize the JADES photometric catalog for the GOODS-South field \citep{rieke2023,eisenstein2023}, which provides flux measurements in both JWST NIRCam and HST ACS \& WFC3 IR bands. We have also employed far-infrared photometry acquired with the Herschel PACS 70, 100, and 160 $\mu$m bands acquired from the GOODS-Herschel catalog \citep{elbaz2011}. To confirm samples with reliable redshift measurements within $z\sim$0.5--0.8, we utilize 3DHST spectroscopic redshift measurements ($z_{\rm spec}$ or $z_{\rm grism}$ with a reliable flag) provided by \citealt{momcheva2016}. To probe the ISM ionizing conditions and gas-phase metallicity of our sample galaxies, we further utilize the MUSE emission line catalog of the CANDELS/GOODS-S field \citep{urrutia2019}, which partly overlaps with the JWST MIRI coverage, as shown in Figure \ref{fig:ds9}.

\begin{table*}
\centering
\caption{Photometric wavebands used for the CIGALE SED analysis of galaxies}
\label{table_bands}
\begin{tabular}{p{1.5cm}p{2cm}p{9cm}p{4.5cm}}
\hline
Telescope & Instrument & Filters & Catalog source\\
 (1) & (2) & (3) & (4) \\\hline
 AstroSat & UVIT & N242W & \citet{saha2024}\\\hline
 KPNO 4m & MOSAIC & U & \citet{skelton2014}\\\hline
 HST & WFC3 UVIS & F275W, F336W & \citet{oesch2018}\\
     & ACS & F435W, F606W, F775W, F814W, F850LP & \citet{rieke2023}\\
     & WFC3 IR & F105W, F125W, F140W, F160W  & \citet{rieke2023} \\\hline
 JWST & NIRCAM & F090W, F115W, F150W, F182M, F200W, F210M, F277W, F335M, F356W, F410M, F430M, F444W, F460M, F480M & \citet{rieke2023} \\
  & MIRI & F560W, F770W, F1000W, F1280W, F1500W, F1800W, F2100W, F2550W & \citet{alberts2024} \\\hline
  Herschel & PACS & PACS blue (70 $\mu$m), PACS green (100 $\mu$m), PACS red (160 $\mu$m) & \citet{elbaz2005}\\
  
\hline
\end{tabular}

\textbf{Note.} Table columns: (1) name of the telescope; (2) name of the instrument; (3) name of the filters that have photometry available; (4) Source of the photometric catalog.
\end{table*}

First, we cross-match the JADES and 3DHST source catalogs to find common sources observed by both JWST and HST within their overlapping area coverage. These sources are further cross-matched with the JWST SMILES source catalog, which resulted in 2819 common sources within the overlapped region of these three surveys. Among the 2819 sources, we find 224 galaxies to have reliable spectroscopic redshifts between 0.5 and 0.8 from the 3DHST catalog (as explained above) and select them as our final samples. The redshift window is fixed based on two conditions: the sensitivity of the defined parameter f$_{\rm PAH}$ (explained in Section \ref{sub_sec:const_fpah} and the effective sampling of the rest-frame FUV emission by the UVIT N242W band. We further cross-match these 224 sources with the AUDF's NUV catalog and find 183 common objects, which represent galaxy samples with detected rest-frame FUV emission. We make sure that none of these UVIT-detected galaxies has a nearby source within 1.2 arcsec (i.e., the full width at half maximum of UVIT's point spread function; \citealt{saha2024,mondal2023}) in the JADES catalog to avoid any possibility of source blending in the UVIT photometry. For the remaining 41 UVIT-undetected sample galaxies, we conduct an additional visual inspection in the UVIT image to confirm the reason for their non-detection. We find 11 galaxies to suffer severe blending for being located near a bright source, while the signal for 30 galaxies falls below the 3$\sigma$ detection limit in the UVIT N242W band.

We notice that 141 of the 224 galaxies are also identified in the Hubble Deep UV Legacy Survey (HDUV) catalog \citep{oesch2018}. We use the HST F275W and F336W band photometry from the HDUV catalog to better constrain the rest-frame UV continuum of these galaxies. 62 of the sample galaxies show detection in at least one or more Herschel PACS band, which helps constrain their FIR continuum better. In Figure \ref{fig:ds9}, we show the spatial distribution of 224 sample galaxies, including the coverage of different surveys mentioned earlier. We also find that 99 of the total galaxies have optical integral field unit (IFU) spectra from the MUSE observations \citep{urrutia2019}. We utilize the MUSE emission line catalog of these 99 galaxies to probe their ISM ionizing condition and gas-phase metallicity. Among 224 sample galaxies, we find 5 to be classified as active galactic nuclei (AGN) in the 7 MS Chandra X-ray source catalog by \citet{luo2017}, and no galaxy is located in the AGN regime of the MIR color-color diagnostics from \citet{donley2012} (Appendix \ref{sec:app_agn}). The number of potential AGNs in our sample becomes 11 when we consider the MIR color-color diagnostics from \citet{lacy2007} (Figure \ref{fig:agn}). As a result, we do not expect AGN to be a significant contaminant to our results.

\begin{figure*}
    \centering
    \includegraphics[width=7in]{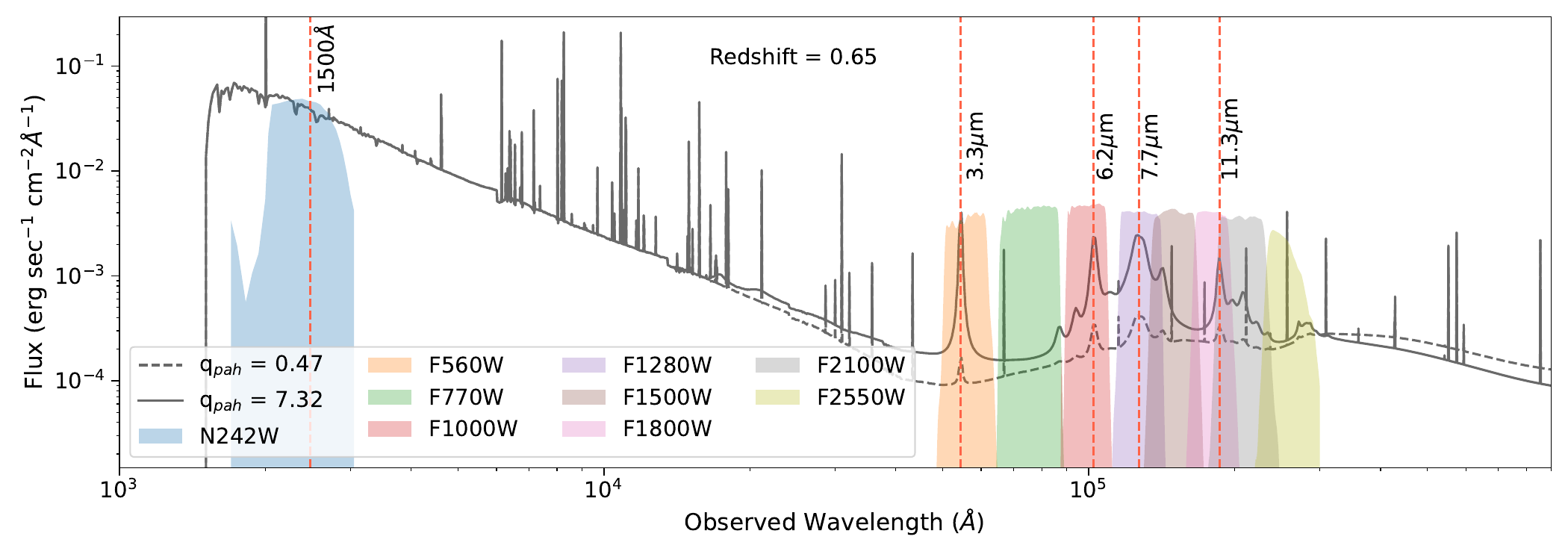}
    \caption{Two CIGALE-generated model spectra, redshifted to $z = 0.65$, representing the lowest (\qpah = 0.47) and highest (\qpah = 7.32) PAH mass fraction values are shown in dashed and solid gray lines, respectively. Four prominent PAH emission bands, at rest-frame wavelengths 3.3, 6.2, 7.7, and 11.3 $\mu$m, are indicated using vertical dashed red lines. The transmission curves of UVIT N242W and eight MIRI filters are also shown to highlight the spectral region covered by each of them. The rest-frame wavelength of 1500 $\AA{}$, marked in a dashed red line, is covered by the UVIT N242W band. Both flux and effective area values are scaled arbitrarily. The figure demonstrates that the parameter f$_{\rm PAH}$, as defined in equation \ref{eq:fpah}, can probe the strength of 7.7$\mu$m PAH emission, and the UVIT N242W filter can sample the rest-frame FUV flux of galaxies at our targeted redshift window.}
    \label{fig:pah_filters}
\end{figure*}

\section{Analysis}
\label{sec:analysis}
\subsection{Model diagnostics} 
\label{sec:model}

\subsubsection{Constructing the parameter f$_{\rm PAH}$}
\label{sub_sec:const_fpah}
To probe the strength of different PAH emission bands in our sample galaxies from the observed MIRI fluxes, we first produce diagnostic relations using model template spectra. We utilize \texttt{dl2014} dust models \citep{dale2014} available within the CIGALE \citep{boquien2019} spectral energy distribution (SED) modeling package to generate galaxy spectra for 11 different values of PAH mass fraction, i.e., \qpah, ranging between 0.46 and 7.32 (Table \ref{table_cigale}). The \qpah value, which denotes the percentage of mass contributed by PAH to the total dust mass, is reported to show a positive correlation with the strength of PAH emission bands in galaxies (e.g., \citealt{shivaei2024}). To bring out the effect of \qpah variation on the PAH emission using template spectra, we fix other CIGALE parameters to optimal values (e.g., age = 6.5 Gyr, burst age = 10 Myr, Metallicity Z = 0.008, log U = $-$3.0, and E(B$-$V)$_{\rm line}$ = 0.3) considering the typical nature of galaxies at $0.5 \leq z \leq 0.8$. In Section \ref{sub_sec:fpah_other_param}, we discuss how the variation of these parameters can affect \qpah. The \texttt{dl2014} model also allows the use of a variable radiation field exposed to the dust in terms of three free parameters: U$_{\rm min}$ - the minimum value of star-light intensity relative to the local interstellar radiation field, $\alpha$ - the power-law index that decides the shape of the radiation field between U$_{\rm min}$ and U$_{\rm max}$, and $\gamma$ - the fraction of dust mass illuminated by star-light. To understand the effect of \qpah variation on PAH emission features, we also fix these three parameters to U$_{\rm min}$ = 10.0, $\alpha$ = 2.0, $\gamma$ = 0.02. In Figure \ref{fig:pah_filters}, we show two such model spectra produced with the lowest (\qpah = 0.46) and highest (\qpah = 7.32) PAH mass fractions. The increasing \qpah value boosts the MIR continuum (within $\sim3 - 12 \mu$m) and also makes the PAH emission features stronger. 

We find that the F1280W filter effectively covers the 7.7$\mu$m broadest PAH feature within the redshift range 0.5 - 0.8, while F770W and F2550W filters sample the MIR continuum on either side of the 7.7$\mu$m emission band (Figure \ref{fig:pah_filters}). Considering this alignment, we define a quantity f$_{\rm PAH}$ (shown in equation \ref{eq:fpah}) as the ratio of flux in the emission band (F1280W) to the average flux in two continuum bands (F770W \& F2550W) to quantify the strength of the 7.7$\mu$m PAH band, which is expected to correlate with the PAH mass fraction (i.e., \qpah). 
\begin{equation}
\label{eq:fpah}
    \rm f_{PAH} = \frac{Flux_{F1280W}}{0.5\times(Flux_{F770W} + Flux_{F2550W})}
\end{equation}
By definition, the parameter f$_{\rm PAH}$ probes the equivalent width of the 7.7$\mu$m PAH feature, which is an important quantity to understand PAH mass fraction as well as PAH excitation in galaxies. Similar attempts using observations in different JWST and Spitzer filter combinations have also been made to probe PAH emission in both nearby \citep{engelbracht2005,khramtsova2014,sandstrom2023,egorov2023,chastenet2023,donnelly2025} and distant \citep{shivaei2024,lin2024,benotto2026} galaxies. Using physical model spectra, \citet{draine2021} showed that the PAH mass fraction can be constrained from the strength of the 7.7$\mu$m emission band, as we attempt to do in this study.

\begin{figure}
    \centering
    \includegraphics[width=3.5in]{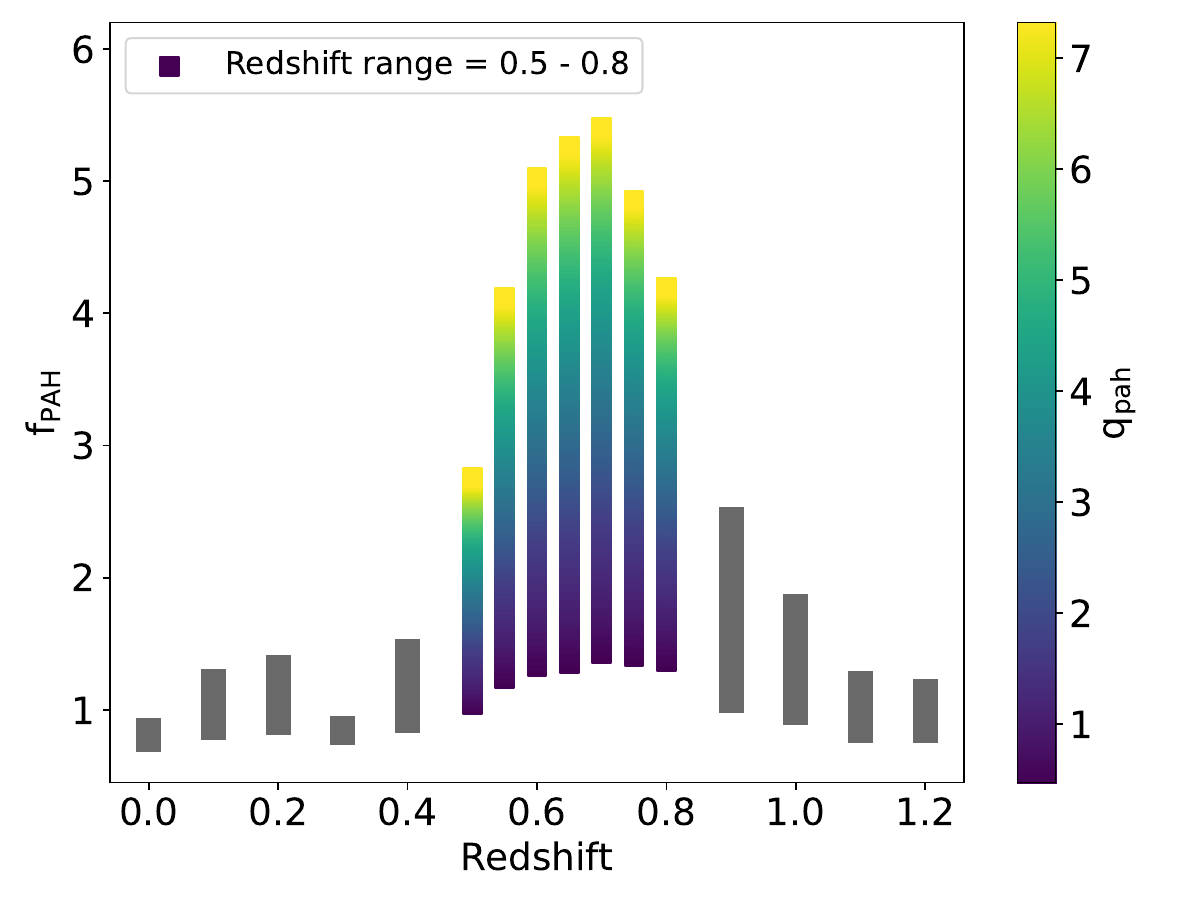}
    \caption{Variation of f$_{\rm PAH}$ with \qpah at different redshifts between 0 and 1.2 as constructed using CIGALE model spectra. The colored bars indicate the redshift range 0.5 -- 0.8, considered in this study, where the colorbar represents \qpah values within the adopted range from 0.47 to 7.32 (listed in Table \ref{table_cigale}). This diagnostic diagram illustrates that the variation of f$_{\rm PAH}$ is most sensitive within our targeted redshift range and hence can be used to probe the strength of 7.7$\mu$m PAH emission in galaxies.}
    \label{fig:fpah_redshift_model}
\end{figure}

To understand the relation between \qpah and f$_{\rm PAH}$, we estimate the expected flux in each MIRI waveband by convolving its effective area profile with the model spectra redshifted to $z=0.65$, which is the mean redshift of our targeted window. Using these fluxes, we compute f$_{\rm PAH}$ for 11 different \qpah values. As a change in the redshift value would shift the observed PAH emission bands with respect to the wavelength span of the MIRI filters, we compute the f$_{\rm PAH}$ values considering 16 different redshift values within 0 -- 1.2. In Figure \ref{fig:fpah_redshift_model}, we show the variation of f$_{\rm PAH}$ for a change in \qpah between 0.47 and 7.32 at different redshifts. The figure shows that the change in f$_{\rm PAH}$ is significant within the redshift range 0.5 - 0.8, whereas it is less sensitive outside this redshift window, where the 7.7$\mu$m emission band moves out of the F1280W waveband. In addition, at redshifts $z<0.5$ or $z>0.8$, both F770W and F2550W continuum bands sample a spectral region consisting of other PAH bands, which further suppresses the defined f$_{\rm PAH}$ value. We note here that the E(B$-$V) value (i.e., 0.3) adopted to produce Figure \ref{fig:fpah_redshift_model}, represents the average value observed in our sample galaxies (with 98\% of galaxies showing E(B$-$V) $<1.0$). This further confirms that our samples do not include heavily dust obscured galaxies for which the MIR continuum shape can deviate substantially from the template spectra used to calibrate f$_{\rm PAH}$. To summarize, the quantity f$_{\rm PAH}$, which we defined to trace the strength of the 7.7$\mu$m PAH band, can efficiently probe the variation of PAH mass fraction in galaxies located within redshift 0.5 -- 0.8. We further note that the f$_{\rm PAH}$ variation is most sensitive within $z = 0.6 - 0.7$, where 38\% of our galaxy samples lie.

\begin{table}
\caption{CIGALE input parameter values used for the SED modeling of the sample galaxies}
\label{table_cigale}

\begin{tabular}{p{2.8cm}p{4.8cm}}
\hline
Parameters & Values\\\hline
\multicolumn{2}{c}{SFH - sfh2exp}\\\hline

Age (Myr) & 1000, 2000, 3000, 4000, 5000, 6000, 7000, 8400\\
$\tau_{main}$ (Myr) & 500, 1000, 2000, 4000, 6000\\
Burst$_{age}$ (Myr) & 20, 100, 250\\
$\tau_{burst}$ (Myr) & 10, 50\\
$f_{burst}$ & 0.05, 0.1\\\hline

\multicolumn{2}{c}{SSP model - BC03 \citep{bruzual2003}}\\\hline

Stellar IMF & Chabrier \citep{chabrier2003}\\
Metallicity & 0.004, 0.008, 0.02\\\hline

\multicolumn{2}{c}{Nebular}\\\hline
logU & $-$3.0\\
f$_{\rm esc}$ & 0\\\hline

\multicolumn{2}{c}{Dust attenuation law}\\\hline
Dust law & Modified Calzetti 2000 \citep{calzetti2000}\\
E(B$-$V)$_{line}$ & 0.01, 0.02, 0.03, 0.05, 0.1, 0.15, 0.2, 0.3, 0.4, 0.5, 0.6, 0.8, 1.0, 1.2, 1.5, 1.8, 2.0\\
E(B$-$V) factor & 0.44\\
Power law slope ($\delta$) & $-$0.7,$-$0.5,$-$0.3, 0.0, 0.3\\\hline

\multicolumn{2}{c}{Dust emission}\\\hline
Dust grain model & dl2014 \citep{dale2014}\\
\qpah & 0.47, 1.12, 1.77, 2.50, 3.19, 3.90, 4.58, 5.26, 5.95, 6.63, 7.32\\
U$_{\rm min}$ & 10\\
$\alpha$ & 2\\
$\gamma$ & 0.02\\\hline

\end{tabular}
\end{table}

\begin{figure}
    \centering
    \includegraphics[width=3.5in]{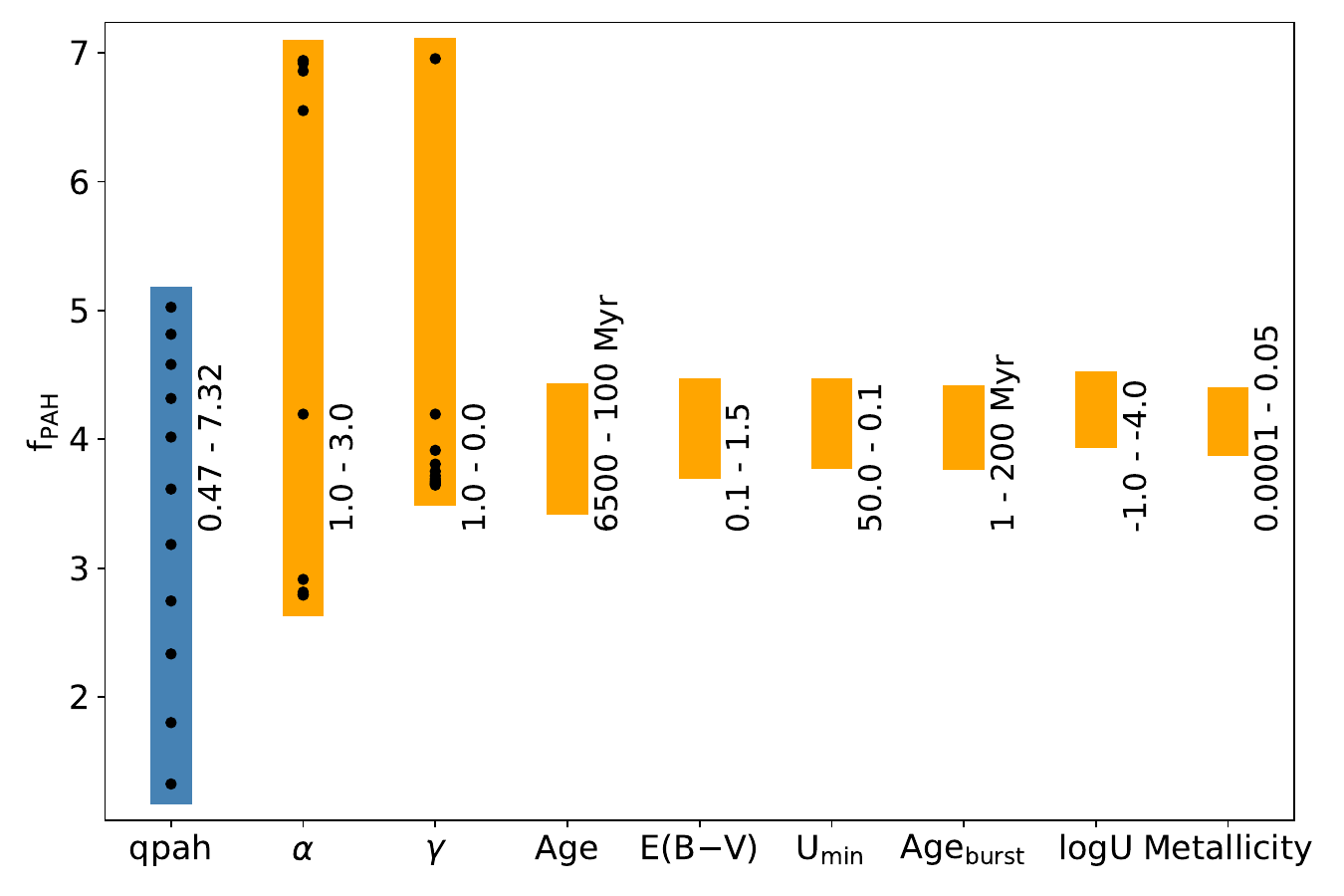}
    \caption{Variation of f$_{\rm PAH}$ at $z =$ 0.65 for the change of different parameters (keeping \qpah = 4.58), which are kept constant to produce Figure \ref{fig:fpah_redshift_model}, are shown in orange bars. The lower and upper limit of each parameter's value is noted, corresponding to the f$_{\rm PAH}$ variation from lower to higher values. The blue bar highlights the range of f$_{\rm PAH}$ (with the black points showing the exact values) for the variation of \qpah at $z = $0.65 as shown in Figure \ref{fig:fpah_redshift_model}. The exact values of f$_{\rm PAH}$ for the variation in $\alpha$ (range: 1 - 3) and $\gamma$ (range: 0 - 1) parameters are also shown using black points on the corresponding bars. The parameter f$_{\rm PAH}$ does not show significant variation for changes in most of the CIGALE input parameters except the dust parameters $q_{\rm PAH}$, $\alpha$, and $\gamma$.}
    \label{fig:fpah_parameters_model}
\end{figure}

\subsubsection{Variation of f$_{\rm PAH}$ with other SED parameters}
\label{sub_sec:fpah_other_param}

Though we have demonstrated f$_{\rm PAH}$ as a good indicator of PAH mass fraction, it is important to check its dependency on other CIGALE parameters that we fixed in our earlier exercise. Considering a fixed value of \qpah (i.e., \qpah=4.58), we produce multiple template sets, each consisting of 11 spectra, by uniformly varying each of the following parameters within the noted range: main stellar population age (range: 100 - 6500 Myr), Burst age (range: 1 - 200 Myr), reddening E(B$-$V) (range: 0.1 - 1.5), ionization parameter log U (range: $-4.0$ - $-1.0$), U$_{\rm min}$ (range: 0.1 - 50.0), $\alpha$ (range: $1$ - $3$), and $\gamma$ (range: 0 - 1). For the metallicity variation, we produce 6 spectra (Z = 0.0001 -- 0.05) as allowed by CIGALE. For each set of model spectra (redshifted to $z=0.65$), we compute f$_{\rm PAH}$ following the same approach to determine its sensitivity to the variation of these parameters at a fixed \qpah value. As shown in Figure \ref{fig:fpah_parameters_model}, apart from the parameters $\alpha$ and $\gamma$, we see insignificant variation of f$_{\rm PAH}$, suggesting f$_{\rm PAH}$ to be a robust parameter for probing 7.7$\mu$m PAH emission strength in a systematically selected galaxy sample. The value of $\alpha$, which determines the shape of the interstellar radiation field between the adopted minimum and maximum limits, has been considered to be 2 in many studies (eg, \citealt{donevski2020,yang2022,chastenet2025}). In constraining the dust parameters of 62 nearby galaxies, \citet{aniano2020} adopted an $\alpha$ range of 1–3, obtaining a median value of 1.95. For a smaller value of $\alpha$ (i.e., $<2$), which corresponds to hotter dust, the MIR continuum rises significantly (demonstrated in Figure \ref{fig:alpha_gamma} in Appendix \ref{app_sed}), resulting in a lower f$_{\rm PAH}$ value. However, a galaxy with intrinsic $\alpha$ greater than 2 can shift the f$_{\rm PAH}$ to an even larger value. We notice no significant change in f$_{\rm PAH}$ for the variation of $\gamma$, which denotes the mass fraction of dust illuminated by emission from stars, above 0. For $\gamma = 0$, we notice the MIR continuum drops significantly (demonstrated in Figure \ref{fig:alpha_gamma}), which in turn boosts the defined f$_{\rm PAH}$ parameter. To summarize, the parameter f$_{\rm PAH}$, defined using fluxes in 3 MIRI filters, can be used to constrain the PAH mass fraction in galaxies within redshift $z = 0.5 - 0.8$, supported by the assumption that $\gamma$ has a non-zero value and $\alpha$ does not vary much across the sample galaxies.

\begin{figure*}
    \centering
    \includegraphics[width=6.5in]{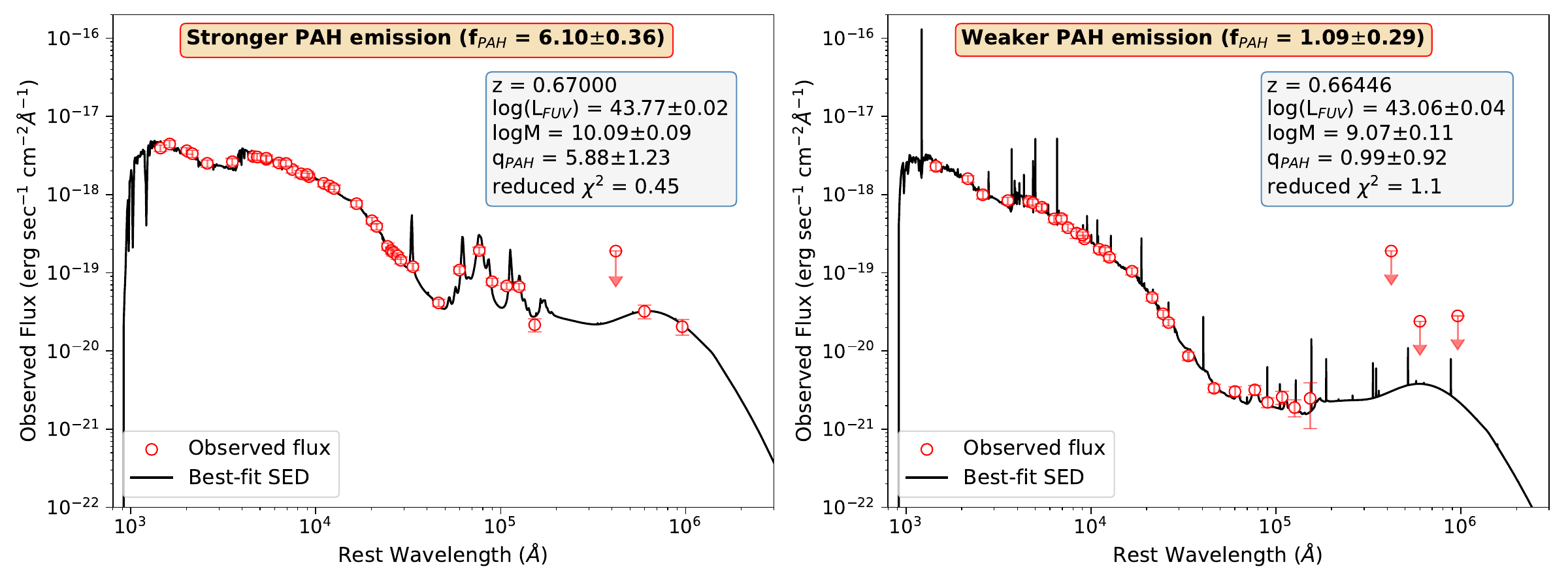}
    \caption{The observed flux and CIGALE best-fit SEDs of two selected galaxies, one with a higher (left) and another with a lower (right) PAH mass fraction. The corresponding values of redshift, f$_{\rm PAH}$, $q_{\rm PAH}$, rest-frame FUV luminosity (log~(L$_{\rm FUV}$)[erg~s$^{-1}$]), stellar mass (log(M/M$_{\odot}$)), and reduced $\chi^2$ are mentioned in respective panels. These best-fit SEDs highlight the robustness of our method to constrain PAH mass fraction using f$_{\rm PAH}$.}
    \label{fig:sed}
\end{figure*}

\begin{figure}
    \centering
    \includegraphics[width=3.5in]{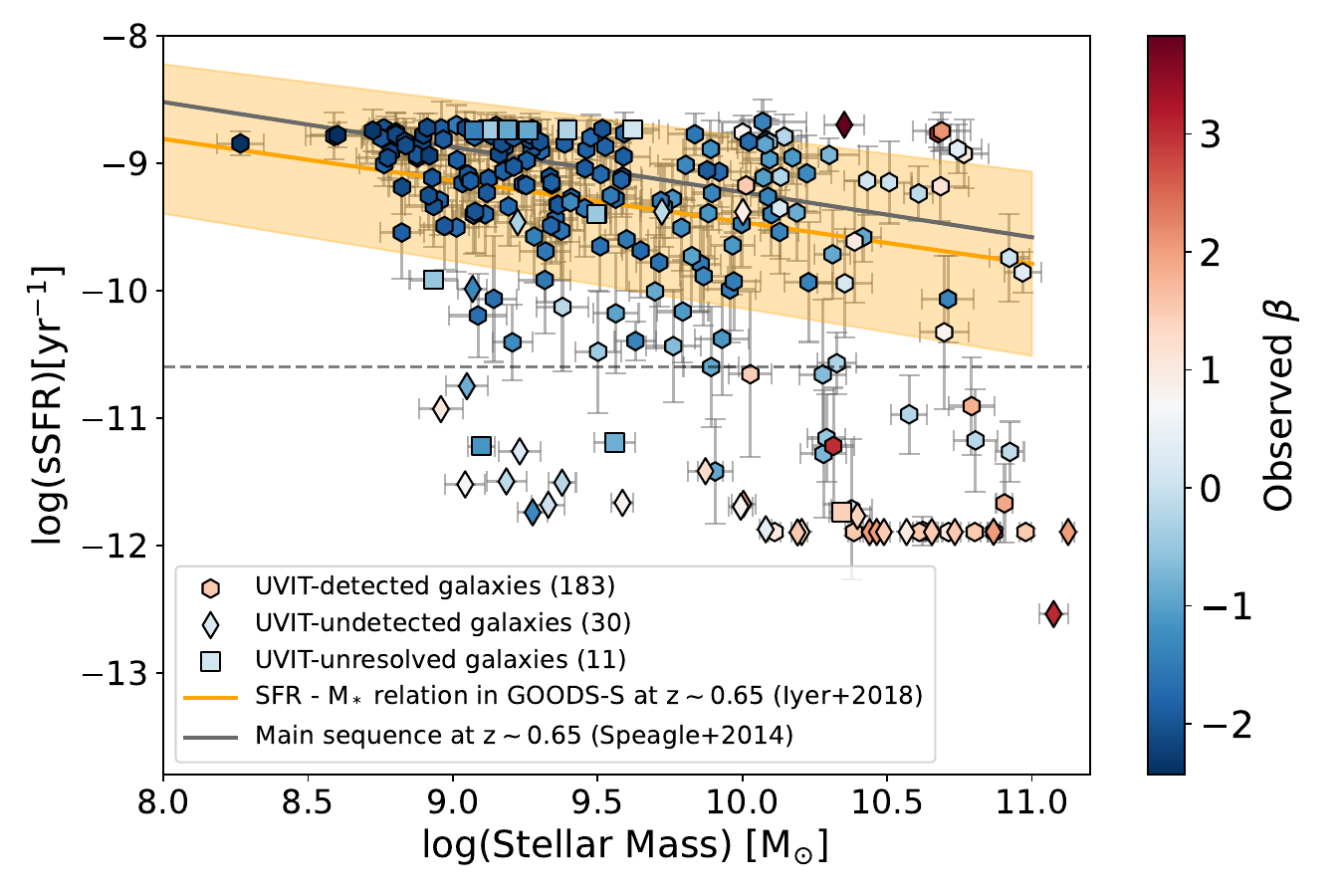}
    \caption{The distribution of stellar mass and specific SFR (over a timescale of 100 Myr) of the 224 sample galaxies (located within redshift 0.5 - 0.8), with the colorbar showing their observed UV continuum slope $\beta$. The values of these three parameters for all sample galaxies are derived from the CIGALE SED modeling. The sSFR-stellar mass relation of galaxies in the GOODS-S field at redshift 0.65 from \citet{iyer2018} is shown using a solid orange line. The shaded region denotes 1$\sigma$ scatter in the relation. We also show the galaxy main sequence at $z\sim0.65$ from \citet{speagle2014} using a solid grey line. Galaxies lying below the horizontal dashed line (i.e., log(sSFR) [yr$^{-1}$] $<-$10.6; the threshold sSFR between SFG and QG at z$\sim$0.65 from \citealt{pacifici2016}) are considered as quiescent galaxies.}
    \label{fig:ms}
\end{figure}

\subsection{CIGALE SED analysis}
\label{sub_sec:cigale_sed}

We use photometry in 38 bands (listed in Table \ref{table_bands}) to perform SED analysis of the selected galaxies using CIGALE \citep{boquien2019}. The goal of the SED analysis is to derive an independent measurement of the PAH mass fraction and other physical parameters, e.g., stellar mass, star formation rate, UV continuum slope, etc. As CIGALE input parameters, we use an exponential star formation history added with a late burst of star formation to account for the rest-frame FUV emission. We choose the Chabrier initial mass function \citep{chabrier2003} and use BC03 \citep{bruzual2003} simple stellar population spectra of three different metallicities (Z = 0.004, 0.008, 0.02). All these parameters are listed in Table \ref{table_cigale}. Considering our primary goal, we allow more flexibility in the CIGALE parameters related to the attenuation law and dust emission. We use the modified Calzetti attenuation law \citep{calzetti2000} with a wider range of E(B$-$V) and power-law slope ($\delta$) values (Table \ref{table_cigale}). For constraining the PAH dust emission, we utilize the \texttt{dl2014} dust emission model as explained in Section \ref{sec:model}. The parameter \qpah is varied across the entire allowed range between 0.47 and 7.32. The other three dust parameters U$_{\rm min}$, $\alpha$, and $\gamma$ are kept fixed to bring out the variation of \qpah across the sample galaxies (as explained in Section \ref{sec:model}). We do not include an AGN component in the SED analysis due to the possible lack of hot dust emission in our sample galaxies, which contains only a small fraction of potential AGNs (i.e., $<7\%$; see Appendix \ref{sec:app_agn}).

We implement a batch-mode fitting technique that utilizes the same set of CIGALE input parameters specified in Table \ref{table_cigale} and perform SED modeling considering the number of available flux values from 38 listed bands (Table \ref{table_bands}) and redshift for each respective galaxy. We have used 3$\sigma$ upper limits for filters that fail to detect a source even if it falls within the band's area of coverage. For consistent flux values across different bands having different PSFs, we use photometry derived using Kron apertures from each respective catalog. We correct the observed fluxes for the Milky Way foreground extinction before feeding them to CIGALE. The foreground extinction in each band is derived using the line-of-sight V-band extinction (A$_V$) from the Schlegel Extinction map \citep{schlegel1998} available on the \textit{Infrared Processing and Analysis Center (IPAC) website}\footnote{https://irsa.ipac.caltech.edu/applications/DUST/} and assuming a Fitzpatrick attenuation law \citep{fitzpatrick1999} with R$_V$ = 3.1. The best-fit SED model is derived following a $\chi^2$ minimization method. For each galaxy, the SED analysis provides a best-fit model spectrum along with measurements of different physical parameters. We note here that 80\% of the sample galaxies have reduced $\chi^2<3$, which confirms the robustness of our SED modeling. In Figure \ref{fig:sed}, we show the observed fluxes along with the best-fit SED for two sample galaxies having significantly different f$_{\rm PAH}$ values to highlight the robustness of our method. The SED analysis also computes the best-fit and Bayesian values of each input parameter listed in Table \ref{table_cigale}.

To know the evolutionary stage of the sample galaxies, we show their SED-derived sSFR (derived using SFR over the last 100 Myr) and stellar mass in Figure \ref{fig:ms}. The majority of our sources follow the SFR-stellar mass sequence of galaxies in GOODS-S at $z=0.65$ as reported by \citet{iyer2018}. The distribution of our galaxies with respect to the galaxy main sequence at the same redshift as reported by \citet{speagle2014} further indicates that most of them are star-forming, while some galaxies are quiescent in nature. We notice that 78\% of the total sample galaxies have log(sSFR)[yr$^{-1}$] higher than $-10.6$, the threshold sSFR value between star-forming and quiescent galaxies at $z\sim0.65$ as calculated from the relation provided in \citet{pacifici2016}. We consider these galaxies as a population of star-forming galaxies (SFGs). The remaining fraction of the sample has log(sSFR)[yr$^{-1}$] $<-10.6$, and they belong to the quiescent galaxy (QG) population at our targeted redshift. We note that the threshold log(sSFR)[yr$^{-1}$] varies by only $\sim$ 0.1 dex over the redshift range 0.5 - 0.8, indicating that the classification of SFGs and QGs remains largely unchanged even when a redshift-dependent threshold is adopted. We also find that adopting broader ranges of the SFH parameters (i.e., burst age and burst fraction) increases the derived sSFR by up to $\sim0.3$ dex, primarily for galaxies with log(sSFR) [yr$^{-1}$]$>-9.0$, with a standard deviation of $\sim0.15$ dex in the sSFR difference, indicating that our interpretations based on sSFR are not significantly affected by these parameter choices. In Figure \ref{fig:ms}, we also show 41 sample galaxies that are either not resolved or detected in the UVIT imaging but identified by the HST and JWST with a reliable spectroscopic redshift between 0.5 and 0.8. The majority of these galaxies, missed by the UVIT, have low sSFR (i.e., log(sSFR) [yr$^{-1}$] $<-10.6$), indicating their quiescent phase.

\begin{figure}
    \centering
    \includegraphics[width=3.5in]{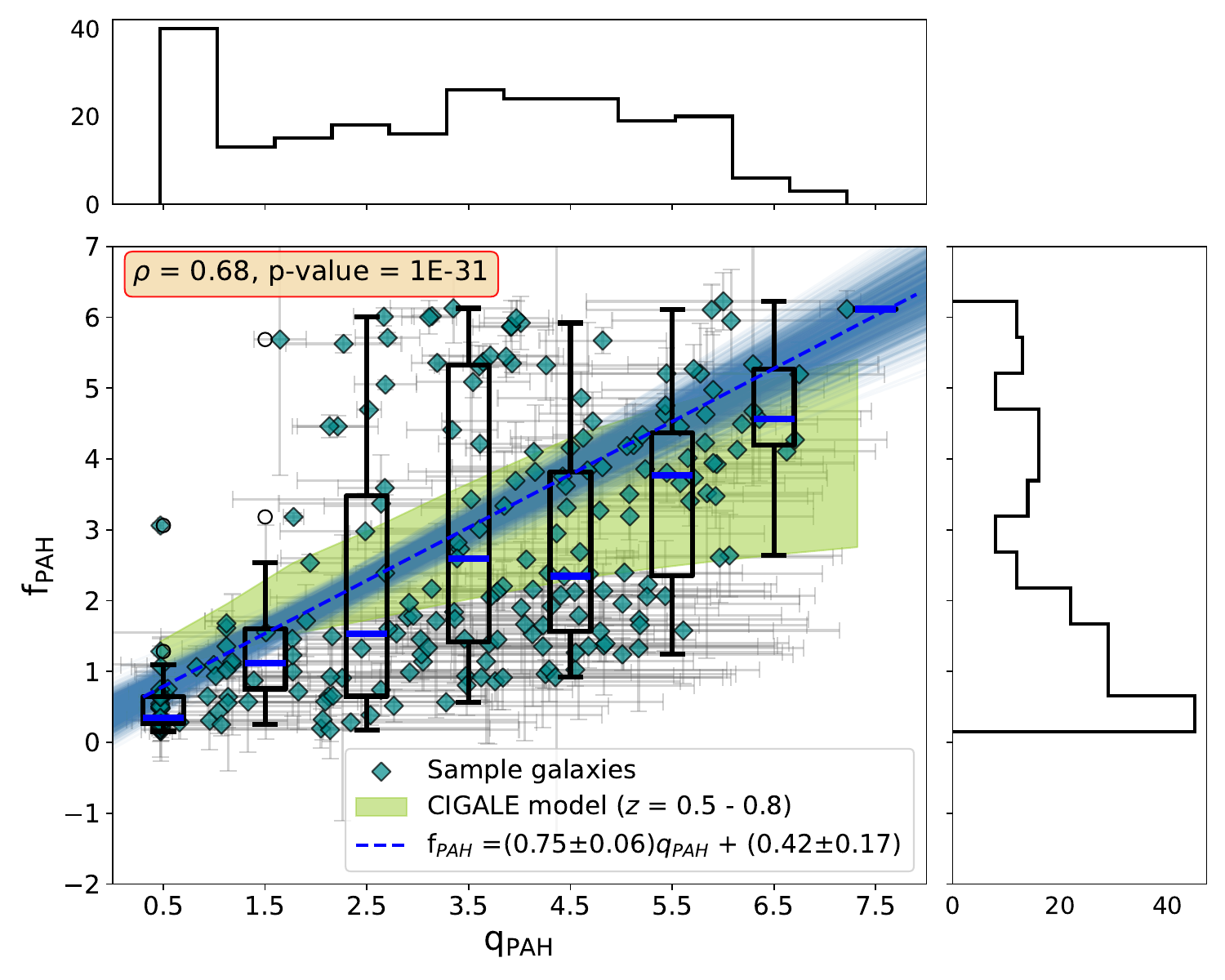}
    
    \caption{The observed value of f$_{\rm PAH}$ and SED-derived Bayesian \qpah of 224 sample galaxies. The distribution is also represented using a box-whisker diagram with seven bins of bin size $\Delta$\qpah = 1.0. The median f$_{\rm PAH}$ in each bin is shown using blue lines, while each box, representing the inter-quartile range (IQR), extends from the first (Q1) to the third (Q3) quartile. The whiskers attached to each box extend up to the farthest data lying beyond the IQR but within 1.5 times the IQR. The blue dashed line (equation \ref{eq_fpah_qpah}) shows the linear relation derived using the hierarchical Bayesian-based linear regression tool \textit{linmix}, suggesting a positive correlation between observed f$_{\rm PAH}$ and model-derived \qpah values. The shaded green region shows the f$_{\rm PAH}$ values covered by synthetic model spectra for \qpah variation from 0.47 to 7.32 within the redshift range 0.5 -- 0.8, as shown in Figure \ref{fig:fpah_redshift_model}. The histograms, adjacent to the top and right axes, show the distribution of the SED-based Bayesian value of \qpah and the observed f$_{\rm PAH}$, respectively.}
    \label{fig:fpah_qpah_box}
\end{figure}

As we have already demonstrated a relation between f$_{\rm PAH}$ and \qpah using model templates in Section \ref{sec:model}, here we use the Bayesian \qpah value of each galaxy measured through the SED modeling and f$_{\rm PAH}$ values measured directly from the extinction-corrected (as explained later in Section \ref{sub_sec:fuv_pah}) MIRI fluxes using equation \ref{eq:fpah} to understand it further. We note here that the f$_{\rm PAH}$ for 41 of the total sample galaxies, which are not detected in the F2550W band, is estimated as the flux ratio of the F1280W to F770W bands. Considering the 5$\sigma$ flux upper limit in the F2550W band, we further derive corresponding lower limits of f$_{\rm PAH}$ for these 41 galaxies. The f$_{\rm PAH}$ lower limits deviate by a small margin from values derived using only F1280W and F770W bands (mean deviation $\sim$ 0.5 with standard deviation $\sim$ 0.4), which assures consistency in our results concerning f$_{\rm PAH}$. In Figure \ref{fig:fpah_qpah_box}, we show our measurements of f$_{\rm PAH}$ and \qpah for all the sample galaxies, along with a box-whisker plot. The uncertainty in \qpah represents the Bayesian uncertainty derived from the SED fitting, while the uncertainty in f$_{\rm PAH}$ is estimated from the flux uncertainties in the three relevant MIRI bands using standard error propagation. We see a positive correlation (Spearman correlation coefficient $\rho$ = 0.68, p-value = 1E-31), signifying that galaxies with increasing \qpah have higher f$_{\rm PAH}$ values, as we expected from our analysis using model templates. We use the \textit{linmix}\footnote{https://linmix.readthedocs.io/en/latest/src/linmix.html} linear regression package to derive a relation between the f$_{\rm PAH}$ and \qpah, which is provided in equation \ref{eq_fpah_qpah}. 
\begin{equation}
\label{eq_fpah_qpah}
    \rm f_{PAH} = (0.75\pm0.06)\times q_{PAH} + (0.42\pm0.17) 
\end{equation} 
 
In Figure \ref{fig:fpah_qpah_box}, we also show the histogram for the model-independent quantity f$_{\rm PAH}$ and the model-derived parameter \qpah to understand their overall distribution. Both histograms show similar trends, with evidence of a bimodal distribution (with peaks at \qpah $\sim$ 0.5 and 4.5, and f$_{\rm PAH}$ $\sim$ 0.3 and 4.2), which further strengthens the robustness of the defined f$_{\rm PAH}$ parameters for probing PAH mass fraction. The observed bimodality could also indicate that PAH enrichment in galaxies goes through a rapid phase while transitioning from a PAH-poor to a PAH-rich system. Considering the correlation between f$_{\rm PAH}$ and \qpah both from the model (Section \ref{sec:model}) and observed data (Section \ref{sub_sec:cigale_sed}), we decide to use the quantity f$_{\rm PAH}$ as an indicator of PAH mass fraction in the following sections, as it is model-independent and easier to derive directly from the observed fluxes. The implementation of f$_{\rm PAH}$ as a probe of PAH mass fraction also demonstrates the uniqueness of JWST/MIRI bands in understanding specific PAH emission bands at higher redshift, where acquiring MIR spectra for a larger galaxy sample is difficult. We further note that substituting SED-based \qpah values for f$_{\rm PAH}$ does not affect our results. We also find that the observed f$_{\rm PAH}$ for our sample galaxies seems to hit an upper ceiling value at $\sim$6.0, which closely agrees with our model-derived value as shown in Figure \ref{fig:fpah_redshift_model}. This could indicate that the PAH mass fraction in galaxies has an upper limit, as has also been assumed in different dust models. Though we find a good correlation, we also see scatter in f$_{\rm PAH}$ values, which is slightly more significant at higher \qpah values. The scatter could arise from multiple factors: photometric error associated with f$_{\rm PAH}$, error in \qpah arising due to the intrinsic variation of different parameters (as discussed in Section \ref{sec:model}) that are assumed constant in our SED modeling, and also due to the redshift variation of selected samples within 0.5 and 0.8 (as demonstrated in Figure \ref{fig:fpah_redshift_model}). In Table \ref{tab:catalog} (Appendix \ref{app_catalog}), we provide the catalog of our sample galaxies that includes their position, redshift, f$_{\rm PAH}$, and other SED-derived parameters.

\begin{figure*}
    \centering
    \includegraphics[width=7in]{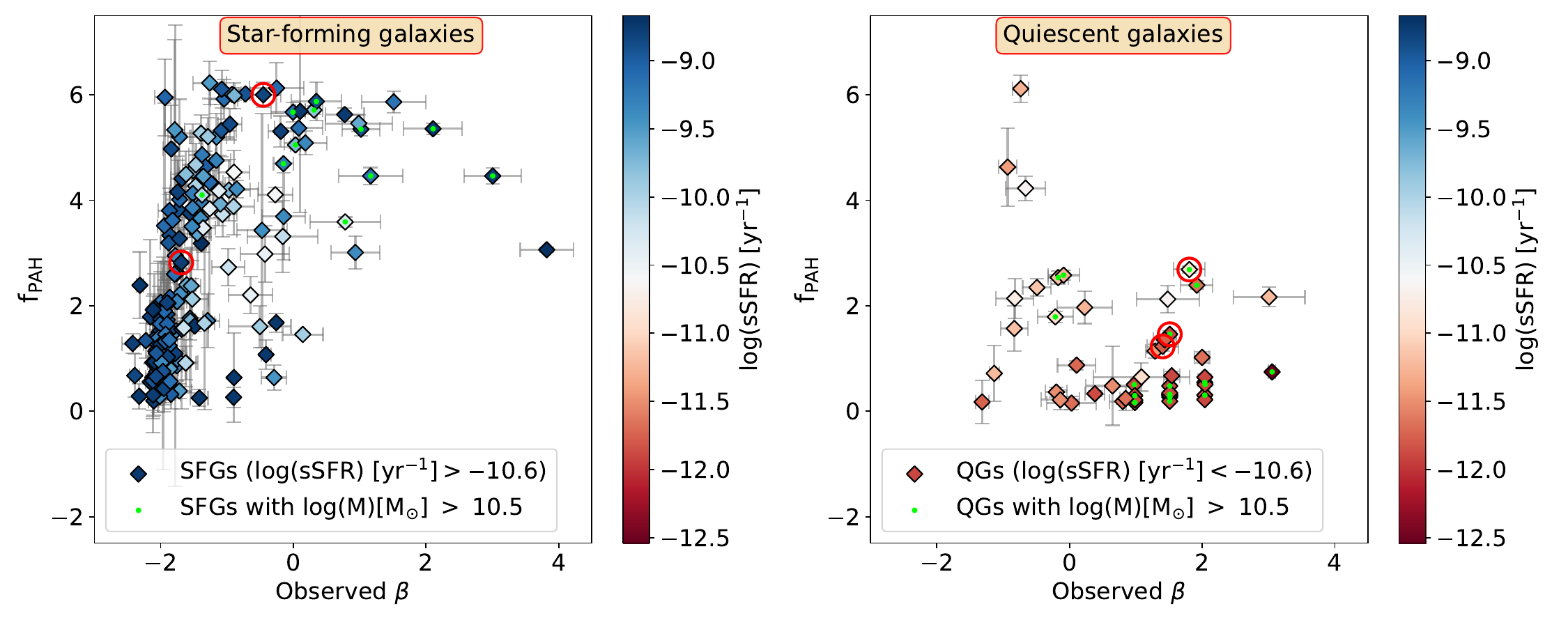}
    \caption{The f$_{\rm PAH}$ and SED-derived observed UV continuum slope $\beta$ of star-forming (left panel) and quiescent (right panel) galaxies with their specific SFR represented using the colorbar. The majority of the SFGs, with log(sSFR)[yr$^{-1}$]$>-10.6$, show a positive correlation between f$_{\rm PAH}$ and observed $\beta$, both of which are sensitive to the dust abundance in galaxies. In the case of quiescent galaxies, we notice a mixed trend, with redder galaxies (i.e., $\beta>0$) especially showing lower f$_{\rm PAH}$ values. In both panels, the most massive galaxies with log(M/M$_{\odot}$)$>$10.5 are marked using green dots. Five of the sample galaxies, confirmed as AGN from the 7 MS Chandra deep survey by \citet{luo2017}, are marked using red circles.}
    \label{fig:fpah_beta_ssfr_mass}
\end{figure*}

\begin{figure}
    \centering
    \includegraphics[width=3.5in]{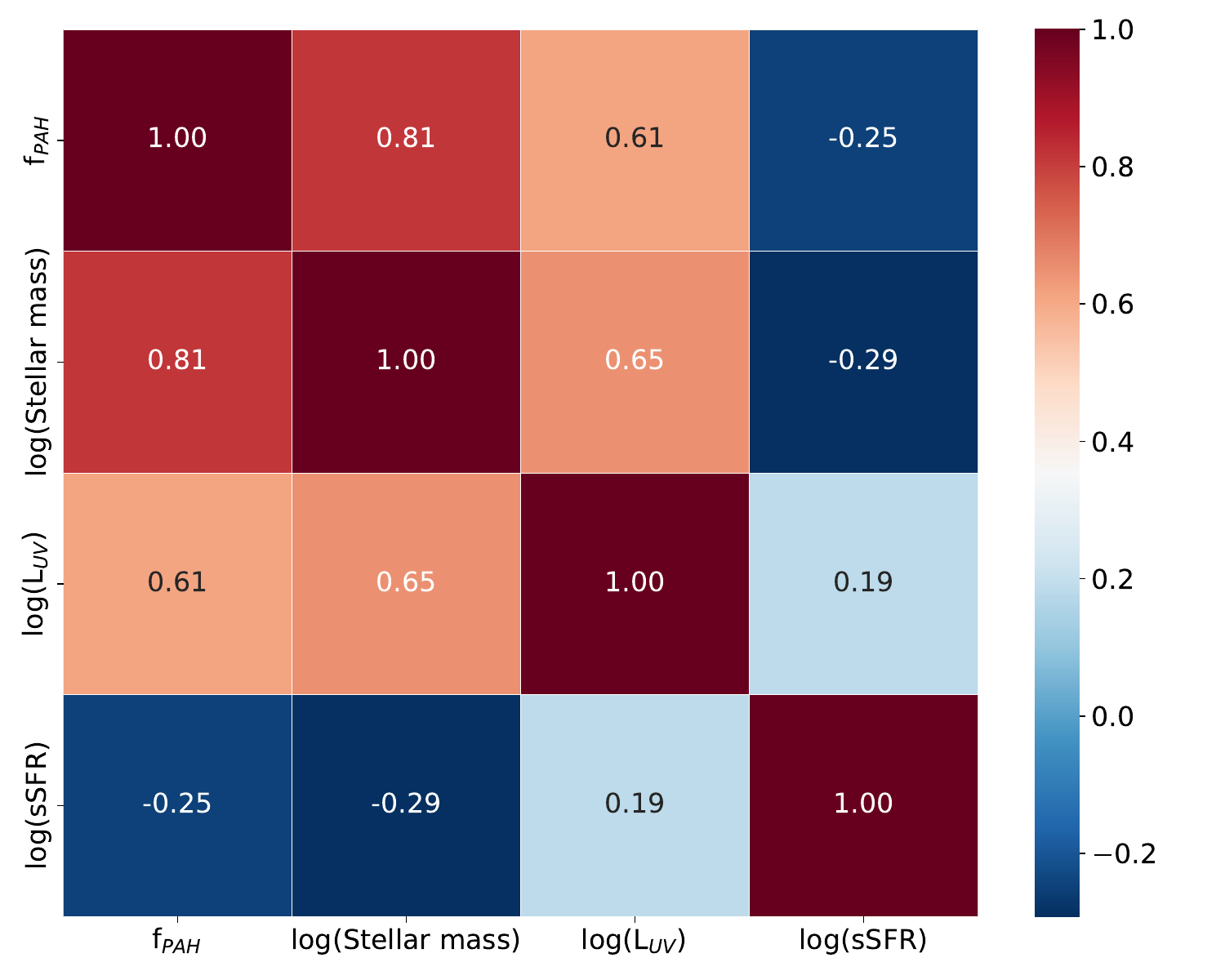}\\
    \caption{Correlation coefficients between f$_{\rm PAH}$ and three measured galaxy parameters (i.e., stellar mass, rest-FUV luminosity, and specific SFR) for 176 SFGs. The galaxy stellar mass shows the tightest correlation with f$_{\rm PAH}$.}
    \label{fig:correlation}
\end{figure}

\section{Results and Discussion}
\label{sec:results_and_discussion}

\subsection{Behavior of PAH emission and observed $\beta$}
\label{sec:pah_beta}

To understand the behavior of f$_{\rm PAH}$ with the total dust extinction in galaxies, we show the estimated values of f$_{\rm PAH}$ and the slope of the observed UV continuum (i.e., $\beta$), as derived from the SED modeling, in Figure \ref{fig:fpah_beta_ssfr_mass}. A higher value of f$_{\rm PAH}$ indicates a larger PAH mass fraction, while the value of observed $\beta$ increases with increasing dust extinction in galaxies \citep{meurer1999,reddy2012}. In Figure \ref{fig:fpah_beta_ssfr_mass} (left panel), we notice a positive correlation between f$_{\rm PAH}$ and $\beta$ (Spearman coefficient = 0.66, p-value = 1E-23) when considering the SFGs. This result indicates that galaxies with higher dust attenuation have a larger PAH mass fraction. However, the value of $\beta$ can also be influenced by the age of stars, where a redder $\beta$ indicates galaxies with older stellar populations and vice versa. The SED-derived age of the main stellar populations in our sample galaxies shows a roughly normal distribution, peaking at 4 Gyr, while 95\% of the galaxies have recent burst populations aged $\geq$100 Myr. Although the age variation may affect the observed $\beta$, we find no clear age - $\beta$ correlation, suggesting that the variation in $\beta$ is primarily driven by differences in internal dust attenuation. Therefore, the correlation between f$_{\rm PAH}$ and $\beta$ could primarily reflect the underlying dependence on stellar mass, given that both the total dust mass and the PAH mass fraction show strong correlations with stellar mass. In other words, our results suggest that as star-forming galaxies build stellar mass with cosmic time, both their total dust content and the fraction of PAH mass with respect to the total dust mass increase. Our observed correlation between f$_{\rm PAH}$ and $\beta$ also aligns well with the relation between dust content and IR emission that has been long-established in the form of the IRX - $\beta$ relation in many studies \citep{calzetti1994,meurer1999,wilkins2013,mondal2024}.

The SFGs with extremely red UV continuum (i.e., $\beta > 0$), the majority of which are massive (with log(M/M$_{\odot}$)$>10.5$ as marked by green dots in Figure \ref{fig:fpah_beta_ssfr_mass}), show higher f$_{\rm PAH}$ values along with a declining trend with increasing $\beta$. As redder $\beta$ would also indicate older stellar populations, these galaxies could plausibly represent a population that is transitioning from a star-forming to a quiescent phase. On the other hand, the selected QGs show a mixed trend between f$_{\rm PAH}$ and observed $\beta$ (Figure \ref{fig:fpah_beta_ssfr_mass} - right panel). The QGs with relatively bluer UV continuum (i.e., $\beta<0$) tend to show a similar correlation observed for SFGs, whereas those with redder UV continuum (i.e., $\beta>0$) are mostly massive (i.e., log(M/M$_{\odot}$)$>10.5$) and show lower f$_{\rm PAH}$. Using JWST/MIRI observations, \citet{blanquez2023} also reported such diversity in MIR emission from a sample of mass-limited QGs at $z\sim0.2-0.7$. The smaller f$_{\rm PAH}$ observed in massive and red QGs could be due to less effective heating of PAH molecules in a less active environment of QGs \citep{egorov2025} or preferential destruction of smaller dust grains through different feedback processes during the last starburst phase in these galaxies \citep{smercina2018}. Our results bring out the contrasting behavior of star-forming and quiescent galaxies in PAH emission with respect to their observed $\beta$. We plan to conduct a detailed study to understand the PAH emission of QGs in a follow-up paper.

\subsection{PAH emission with varying galaxy properties}
\label{sub_sec:pah_galaxy_properties}
The PAH emission in galaxies depends on two primary factors: the fraction of PAH mass compared to the total dust (i.e., \qpah) and the amount of available UV photons that can excite the PAH molecules \citep{draine2021}. The parameter \qpah is known to scale with the galaxy stellar and gas mass \citep{cortzen2019,shivaei2024}, whereas the rate of recent or ongoing star formation primarily decides the UV luminosity of a galaxy. In Figure \ref{fig:correlation}, we show the correlation coefficients calculated among four parameters, i.e., f$_{\rm PAH}$, stellar mass, rest-frame FUV luminosity, and specific SFR for 176 SFGs. The parameter f$_{\rm PAH}$ shows the tightest correlation with galaxy stellar mass ($\rho$ = 0.81), which also strongly correlates with the FUV luminosity. On the other hand, the specific SFR shows a weak anti-correlation with f$_{\rm PAH}$, suggesting plausible suppression of PAH abundance in galaxies with higher sSFR. We discuss these relations in detail in the following sections.

\begin{figure*}
    \centering
    \includegraphics[width=7in]{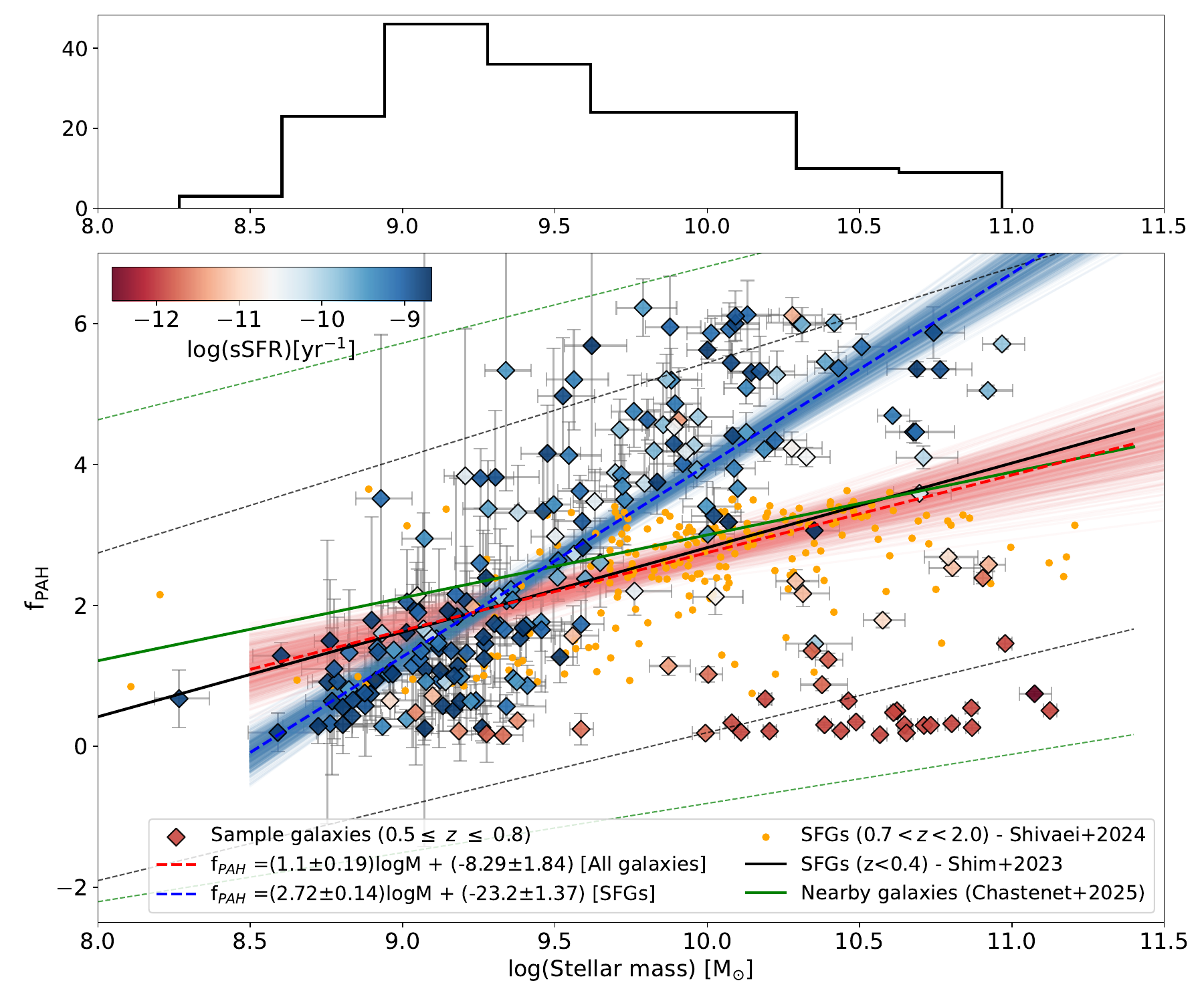}
    \caption{The SED-derived stellar mass and observed f$_{\rm PAH}$ of all sample galaxies. The specific SFR values are indicated using a colorbar. The dashed red line shows the linear relation between f$_{\rm PAH}$ and stellar mass derived using \textit{linmix} package for the log(stellar mass) range 8.5 - 11.0. The SFGs with log(sSFR)[yr$^{-1}$]$>-10.6$ show an even stronger positive correlation ($\rho$ = 0.81) and the corresponding linear relation is shown using the dashed blue line. The relation between \qpah and stellar mass from other studies is shown (SFGs at $z = 0.7 - 2.0$ (orange points) - \citealt{shivaei2024}; SFGs at $z<0.4$ (solid black line) - \citealt{shim2023}; nearby galaxies (solid green line) - \citealt{chastenet2025}) by converting their \qpah to f$_{\rm PAH}$ using equation \ref{eq_fpah_qpah}. The dashed lines show 1$\sigma$ scatter as reported in respective studies. The histogram, adjacent to the top axis, shows the distribution of log(Stellar mass) [M$_{\odot}$] of our sample galaxies.}
   \label{fig:fpah_mass}
\end{figure*}

\subsubsection{Relation with galaxy stellar mass}
\label{sub_sec:fpah_mass}

The mass fraction of PAH is known to show a strong dependence on the stellar mass of galaxies in both the nearby \citep{chastenet2025} and distant universe \citep{shim2023,shivaei2024}. In Figure \ref{fig:fpah_mass}, we show the SED-derived stellar mass and the observed f$_{\rm PAH}$ of all sample galaxies, with their sSFR represented by a colorbar. We notice a positive correlation between f$_{\rm PAH}$ and log(Stellar mass) ($\rho = 0.38$) and show the best-fit linear relation in equation \ref{eq_fpah_mass}.
\begin{equation}
\label{eq_fpah_mass}
    \rm f_{PAH} = (1.10\pm0.19)log(M/M_{\odot}) +(-8.29\pm1.84)
\end{equation}
We find that the selected SFGs follow an even tighter correlation with $\rho = 0.81$ as shown in Figure \ref{fig:correlation}. We further notice that low-mass galaxies with high sSFR have low f$_{\rm PAH}$, which plausibly suggests that star-forming dwarf galaxies have less PAH enrichment. Considering star-forming galaxies at $z = 0.7 - 2.0$ in the same GOODS-S field, \citet{shivaei2024} reported a positive correlation between \qpah and galaxy stellar mass. We show their samples with robust \qpah values in Figure \ref{fig:fpah_mass}. We note here that the \qpah values from other studies are converted to f$_{\rm PAH}$ using equation \ref{eq_fpah_qpah} to show in Figure \ref{fig:fpah_mass} as well as in other figures. Though our results agree well with their measurements in the low-mass range, we find larger f$_{\rm PAH}$ values for massive galaxies. While we allow \qpah to vary between 0.47 and 7.32 (as allowed in CIGALE), \citet{shivaei2024} adopted the upper limit of \qpah as 4.6, which possibly resulted in the different f$_{\rm PAH}$ values that we see in the higher mass range. Considering the same \qpah range as our study, \citet{shim2023} found a f$_{\rm PAH}$-stellar mass relation similar to ours for star-forming galaxies at a relatively lower redshift (i.e., $z<0.4$) (Figure \ref{fig:fpah_mass}). A similar dependence of \qpah on stellar mass is also reported for nearby galaxies by \citet{chastenet2025}. Therefore, our study, targeted at a redshift window not explored well, reinforces galaxy stellar mass as one of the most important factors to decide the PAH mass fraction. As \qpah is known to scale with the amount of metal content in galaxies \citep{chastenet2023,shim2023,whitcomb2024,shivaei2024}, the observed correlation between stellar mass and f$_{\rm PAH}$ could also be driven by metallicity, which increases with galaxy stellar mass \citep{tremonti2004,sanders2021}. We discuss the metallicity dependence of f$_{\rm PAH}$ in Section \ref{sub_sec:metallicity}. 

In addition, the nature of QGs stands out again in the f$_{\rm PAH}$ - stellar mass relation, as also shown in Figure \ref{fig:fpah_beta_ssfr_mass} earlier. The increasing trend of f$_{\rm PAH}$ with stellar mass breaks when the galaxy moves to a quiescent phase, resulting in a much lower PAH mass fraction with respect to star-forming galaxies of similar mass. As noted earlier in section \ref{sec:pah_beta}, the lower f$_{\rm PAH}$ in the majority of massive QGs could be either due to intrinsically lower PAH mass fraction resulting from PAH destruction during a recent quenching episode \citep{smercina2018} or less effective heating of PAH molecules in their passive environment.

\begin{figure*}
    \centering

    \includegraphics[width=7in]{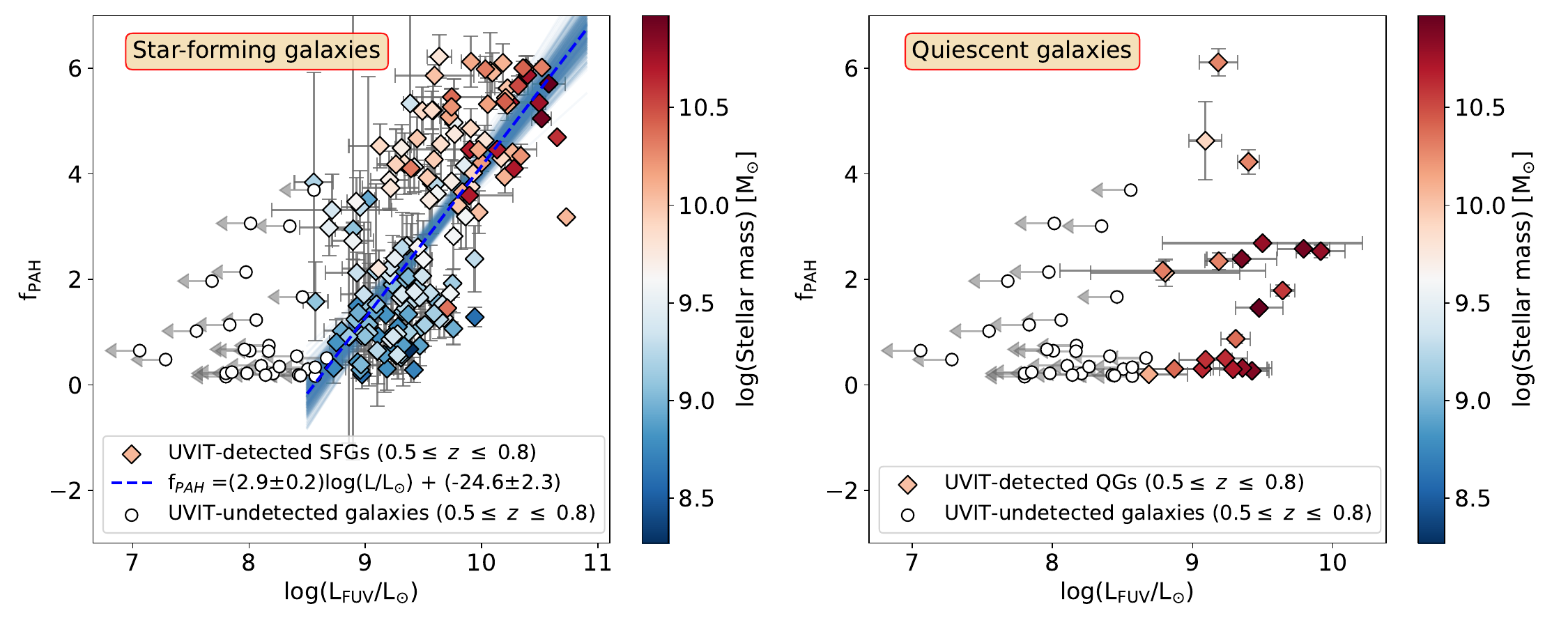}
    \caption{The derived f$_{\rm PAH}$ and extinction-corrected rest-frame FUV luminosity (log(L$_{\rm FUV}$/L$_{\odot}$)) of 183 UVIT-detected sample galaxies (left panel - 162 SFGs, right panel - 21 QGs). The dashed blue line, on the left panel, shows the linear relation obtained using \textit{linmix}. The galaxy stellar mass is shown using the colorbar. The L$_{\rm FUV}$ upper limits from best-fit SEDs of 30 galaxies that are not identified by UVIT but detected in the JWST MIRI bands are shown using open circles. The observed correlation for SFGs signifies that PAH emission strengthens with increasing FUV luminosity. Despite their high stellar masses and moderate FUV luminosities, the UVIT-detected QGs generally show lower f$_{\rm PAH}$ values (right panel).}
    \label{fig:fpah_muv_mass}
\end{figure*}

\subsubsection{Relation with FUV luminosity}
\label{sub_sec:fuv_pah}
FUV emission, mostly contributed by massive young stars, scales with the ongoing SFR of galaxies \citep{kennicutt1998,kennicutt2012}, which is further reported to correlate well with the PAH luminosity in both nearby and distant galaxies \citep{peeters2004,shipley2016,xie2019,lai2020,ronayne2024}. PAH particles are primarily excited by FUV photons, with photon energies of approximately 5.7 - 13.6 eV (which corresponds to $\lambda\sim$ 912 - 2175 $\AA{}$), predominantly generated by stellar populations younger than 100 Myr \citep{draine2021}.
We use the estimated f$_{\rm PAH}$ and extinction-corrected rest-frame FUV luminosity (L$_{\rm FUV}$, i.e., luminosity at rest-frame $\lambda\sim$1500 $\AA{}$) of the UVIT-detected SFGs and QGs (shown in Figure \ref{fig:fpah_muv_mass}) to understand the relation between PAH emission and UV luminosity at $z\sim0.65$. L$_{\rm FUV}$ is calculated from the CIGALE-derived rest-frame luminosity in the N242W band (which effectively samples rest-frame luminosity at $\lambda\sim$1500 $\AA{}$ for $z\sim0.5-0.8$) after correcting it for both the galaxy's internal (using the SED-derived E(B$-$V) values) and Milky Way (as explained in Section \ref{sub_sec:cigale_sed}) extinction. Though a higher value of PAH mass fraction is expected to enhance the strength of PAH emission features, it would also partly depend on the amount of UV photons that are available to interact with the PAH molecules to excite them. Therefore, galaxies with the same \qpah value can result in different PAH emission strengths depending on the difference in their UV radiation field. For UVIT-detected 162 SFGs, we notice a gradual increase in f$_{\rm PAH}$ value towards brighter FUV luminosity, which signifies that a stronger UV emission results in more efficient excitation of the PAH molecules in these galaxies. We perform a linear fit using \textit{linmix} between f$_{\rm PAH}$ and log(L$_{\rm FUV}$/L$_{\odot}$) for these SFGs (Figure \ref{fig:fpah_muv_mass}) and provide the relation in equation \ref{eq:fpah_muv}. 
\begin{equation}
\label{eq:fpah_muv}
    \rm f_{PAH} = (2.9\pm0.2)\times log(L_{\rm FUV}/L_{\odot}) + (-24.6\pm2.3)
\end{equation}

Alternatively, the relation also suggests that SFGs with an increasingly higher rate of ongoing star formation produce a higher amount of PAH molecules, which is reflected in their higher observed f$_{\rm PAH}$. It is equally important to mention that the galaxy stellar mass shows a tight correlation with the L$_{\rm FUV}$ (Figure \ref{fig:fpah_muv_mass} - left panel), which indicates that the observed correlation between f$_{\rm PAH}$ and L$_{\rm FUV}$ could actually be driven by galaxy mass. As galaxy mass correlates with metallicity, massive systems are expected to host ISM conditions that efficiently support PAH production (discussed in Section \ref{sub_sec:metallicity}). To summarize, the massive UV-luminous SFGs are most efficient in producing as well as exciting PAH molecules. However, we also notice significant scatter in f$_{\rm PAH}$ across the covered L$_{\rm FUV}$ range, which appears to be primarily driven by stellar mass. This signature further signifies that stellar mass plays a more fundamental role in driving \qpah in galaxies than FUV luminosity. We also find that most of the QGs detected in the UVIT show relatively lower f$_{\rm PAH}$, despite having moderate UV luminosity and higher stellar mass (Figure \ref{fig:fpah_muv_mass} - right panel). This finding could directly imply that the lower PAH mass fraction observed in QGs is primarily driven by the destruction of these small dust grains through the same physical processes that suppressed star-forming activity in these systems.

\begin{figure}
    \centering
    \includegraphics[width=3.5in]{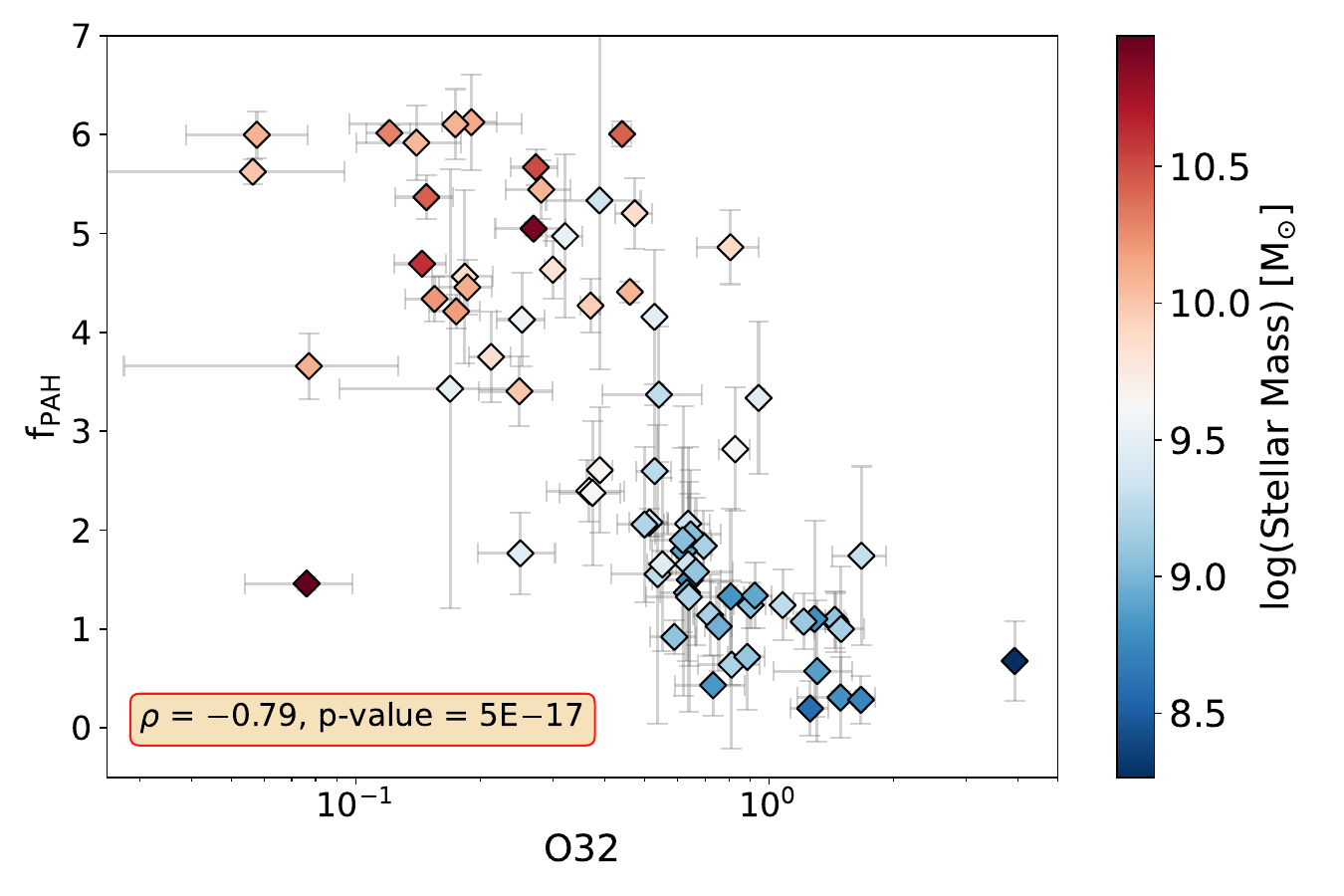}\\
    \includegraphics[width=3.5in]{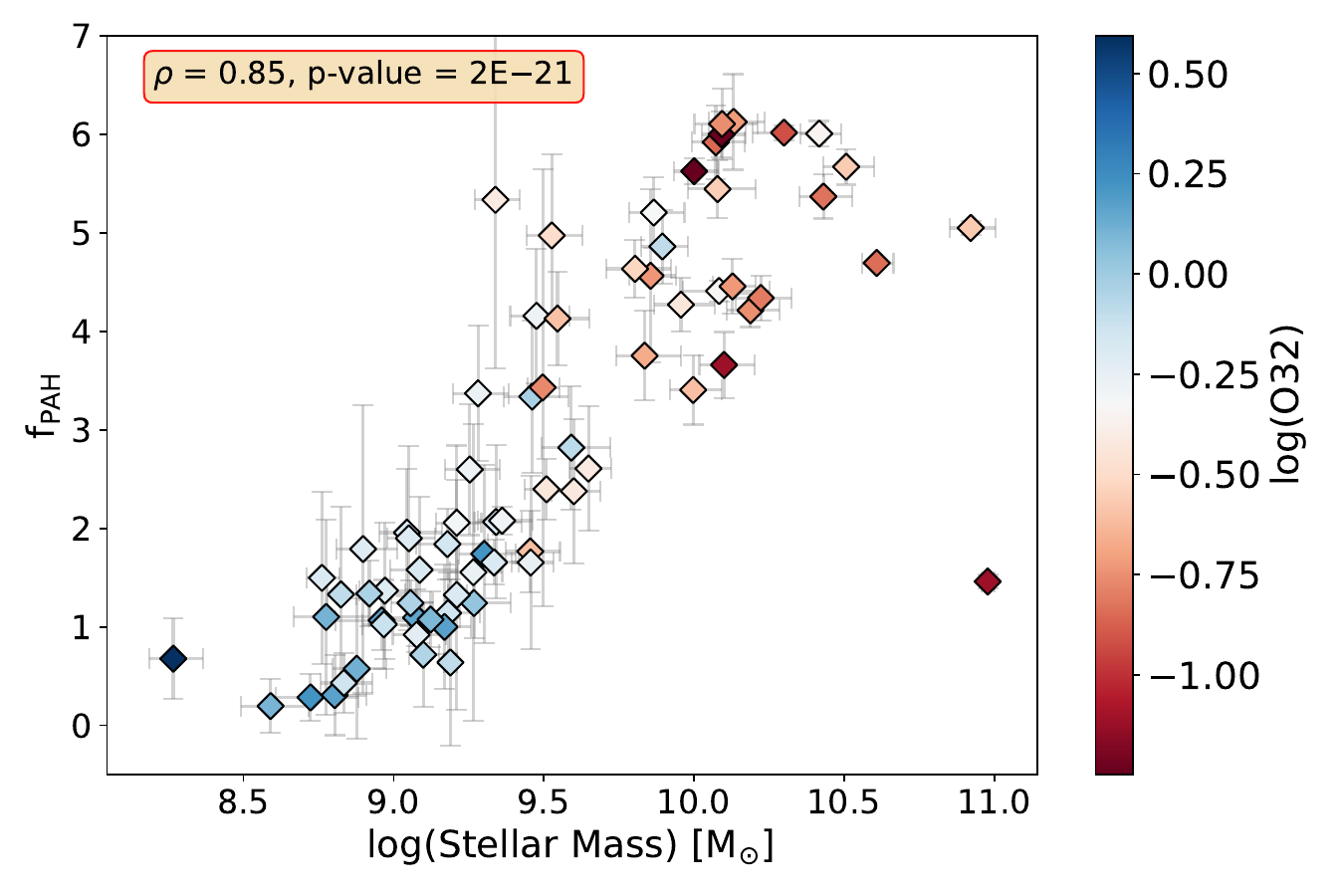}
    \caption{\textbf{Top:} The derived f$_{\rm PAH}$ and O32 (i.e., [OIII]5007/[OII]3727,3729) values of 73 sample galaxies, which are found to show these oxygen lines in their optical MUSE spectra. The colorbar represents the stellar mass of the galaxies. The observed anti-correlation between f$_{\rm PAH}$ and O32 suggests that galaxies with highly ionized ISM tend to have lower PAH content and vice versa. As shown using the colorbar, the stellar mass plays an equally important role in determining the PAH mass fraction in galaxies alongside ISM ionization. \textbf{Bottom:} f$_{\rm PAH}$ and Stellar mass of the same 73 galaxies color-coded with their O32 values. The figure shows a tight correlation between f$_{\rm PAH}$ and stellar mass, which is also shown in Figure \ref{fig:fpah_mass}. It also indicates that, at a fixed stellar mass, galaxies with higher O32 values tend to have lower $f_{\rm PAH}$ (clearer in the low mass regime), further reinforcing the role of harsh ionizing conditions in the photo-dissociation of PAHs.}
    \label{fig:fpah_o32}
\end{figure}

\subsection{PAH emission and ISM ionization}
To further understand how PAH excitation or destruction depends on the interstellar radiation field, we derive the ISM ionization strength of galaxies using optical emission lines observed using the MUSE IFU observations. We find that 99 of the 224 sample galaxies have MUSE CDFS observations obtained by \citet{urrutia2019}. We use the MUSE emission line catalog, produced by \citet{urrutia2019}, to further identify 73 galaxies (71 SFGs and 2 QGs) that show both [OIII]~$\lambda$5007 and [OII]~$\lambda$3727 lines. We use the emission-line fluxes from \citet{urrutia2019} and correct them for both internal and Milky Way extinction (as explained earlier) to derive the line ratio O32, defined as [OIII]~$\lambda$5007/[OII]~$\lambda$3727. The value of O32 is sensitive to the ISM ionization parameter of galaxies \citep{nakajima2014,kewley2019}. A higher O32 value would indicate a strongly ionized ISM, which signifies the production of a larger number of harder UV photons. In Figure \ref{fig:fpah_o32} (top), we show the measured f$_{\rm PAH}$ and O32 values of 73 galaxies, including their stellar mass, using a colorbar. We notice an anti-correlation (Spearman correlation coefficient $\rho = -0.79$) between f$_{\rm PAH}$ and O32, which shows that galaxies with higher O32 have lower f$_{\rm PAH}$ and vice versa. This observed trend implies that galaxies with strongly ionized ISM have a lower PAH mass fraction. In other words, our result suggests that the PAH molecules can effectively be destroyed with increasing ionizing strength of the ISM radiation field in galaxies, which is also reported in both nearby \citep{madden2006,hunt2010,egorov2023,chastenet2023,pedrini2024,egorov2025} and distant \citep{shivaei2017} galaxies. O32 also shows a tighter relation with galaxy stellar mass (Figure \ref{fig:fpah_o32}). All galaxies with a stronger ISM ionizing field (i.e., O32$>$1) have a lower PAH mass fraction and are also less massive, whereas the galaxies with the lowest O32 value are more massive and have a higher PAH mass fraction. To better understand the role of ionization strength, we show f$_{\rm PAH}$ and stellar mass of these 73 galaxies, including O32 as a colorbar in Figure \ref{fig:fpah_o32} (bottom). This sub-sample of galaxies shows an even tighter correlation ($\rho$ = 0.85) between f$_{\rm PAH}$ and stellar mass than what we see for SFGs in Figure \ref{fig:fpah_mass}. Besides, we also find that at a fixed stellar mass, galaxies with weaker ionization strength tend to show relatively higher f$_{\rm PAH}$ values. This behaviour is more evident for the low-mass galaxies with log(M/M$_{\odot}$) $<9.5$ (Figure \ref{fig:fpah_o32} - bottom). While the stellar mass of galaxies can independently play a crucial role in determining the value of f$_{\rm PAH}$, our analysis shows that the ionization conditions of the ISM are equally important in controlling the PAH content in galaxies, particularly in star-forming dwarfs.

\begin{figure}
    \centering
    \includegraphics[width=3.5in]{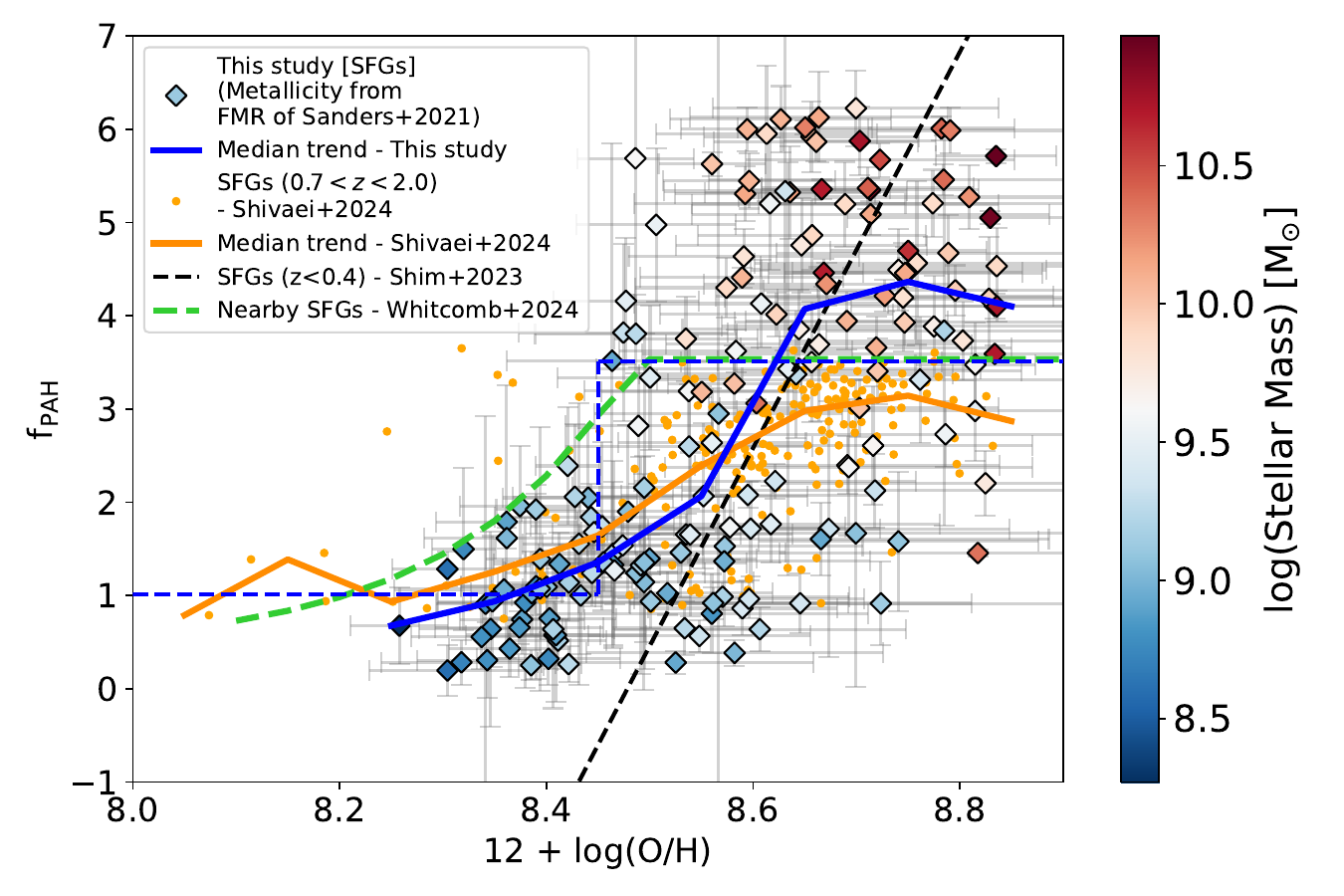}
    \caption{The f$_{\rm PAH}$ and gas-phase metallicity (as derived using FMR from \citealt{sanders2021}) of sample SFGs with stellar mass shown using the colorbar. The moving median for our measurements is shown using a solid blue line. The orange points show SFGs at $z = 0.7 - 2.0$ from \citet{shivaei2024}, and the corresponding moving median is shown in a solid orange line. The f$_{\rm PAH}$ - metallicity relations for nearby \citep{whitcomb2024} and $z<0.4$ \citep{shim2023} SFGs are also shown using dashed green and black lines, respectively. To highlight the threshold metallicity (i.e., 12 + log(O/H) $\sim$ 8.45), below which the PAH mass fraction drops significantly, we also show the step function fit (dashed blue line) to our measurements using the median f$_{\rm PAH}$ of galaxies above and below 12 + log(O/H) = 8.45. The overall positive correlation between f$_{\rm PAH}$ and 12 + log(O/H), as well as f$_{\rm PAH}$ and stellar mass, suggests that massive metal-rich star-forming galaxies have a higher PAH mass fraction than the metal-poor, less massive galaxies.}
    \label{fig:fpah_met}
\end{figure}

\subsection{PAH emission and gas-phase metallicity}
\label{sub_sec:metallicity}
The mass fraction of carbon-based PAH dust molecules is known to correlate with galaxy metallicity \citep{engelbracht2005,draine2007,shivaei2017,shim2023,shivaei2024}. As a result, the metal-rich galaxies are expected to have higher PAH content than the metal-poor ones. Some studies have also shown that galaxies maintain an approximately constant PAH mass fraction above a threshold metallicity, with \qpah decreasing significantly below this threshold \citep{shim2023,shivaei2024,whitcomb2024,aniano2020}. Considering the fundamental metallicity relation (FMR) from \citet{sanders2021}, constructed using star-forming galaxies at $z\sim2.3$ and $z\sim3.3$ (which shows remarkable coincidence with the FMR at $z=0$), we derive the gas-phase metallicity of our selected SFGs from their mass and SFR values. In Figure \ref{fig:fpah_met}, we show the measured metallicity and f$_{\rm PAH}$, which show a positive correlation, indicating PAH enrichment in galaxies with increasing metallicity. \citet{shim2023} reported a similar relation between PAH mass fraction and gas-phase metallicity (which they derived using emission-line metallicity calibrations) for SFGs at z$<0.4$ (Figure \ref{fig:fpah_met}). \citet{whitcomb2024} studied selected regions from three nearby SFGs and reported that the total PAH emission over the total infrared emission remains almost constant at metallicity 12 + log(O/H) $>$ 8.5 but declines significantly below this threshold value (Figure \ref{fig:fpah_met}). For SFGs at $z<0.4$, \citet{shim2023} also adopted 12 + log(O/H) $=$ 8.5 to differentiate galaxies with higher and lower \qpah. For our SFGs, we notice that f$_{\rm PAH}$ starts to drop significantly below the metallicity 12 + log(O/H) $\sim$ 8.45 (i.e., Z $\sim$ 0.7Z$_{\odot}$), which we define as the threshold metallicity for PAH enrichment at $z\sim0.65$ (Figure \ref{fig:fpah_met}). We find the median values of f$_{\rm PAH}$ at 12 + log(O/H) $>8.45$ and $<8.45$ as 1.01 and 3.51, respectively, which closely match the values reported in \citet{whitcomb2024} for nearby SFGs. For SFGs at slightly higher redshift between 0.7 and 2.0, \citet{shivaei2024} used the FMR-based metallicity, derived using the calibration from \citet{sanders2021}, to report a similar trend to ours for the moving median of f$_{\rm PAH}$ (Figure \ref{fig:fpah_met}). In the high-metallicity regime, the $f_{\rm PAH}$ values derived by \citet{shivaei2024} remain consistently lower, possibly owing to the adoption of an upper limit of \qpah = 4.6 in their dust models. Our study therefore indicates that galaxies with 12 + log(O/H)$<$8.45, which are also less massive, have lower PAH mass fraction (f$_{\rm PAH}$ $\lesssim 2.0$), whereas more massive galaxies with metallicities above this threshold exhibit higher PAH abundances. This behaviour of f$_{\rm PAH}$ around the metallicity 12 + log(O/H) $\sim$8.45 could indicate that PAH enrichment expedites above this threshold metallicity in star-forming galaxies at $z\sim0.65$. We further conclude that a threshold metallicity for determining PAH mass fraction does exist for SFGs at $\sim$0.65, and its close agreement with the threshold values reported at other redshifts, including $z = 0$, suggests minimal evolution with cosmic time.

\begin{figure}
    \centering
    \includegraphics[width=3.5in]{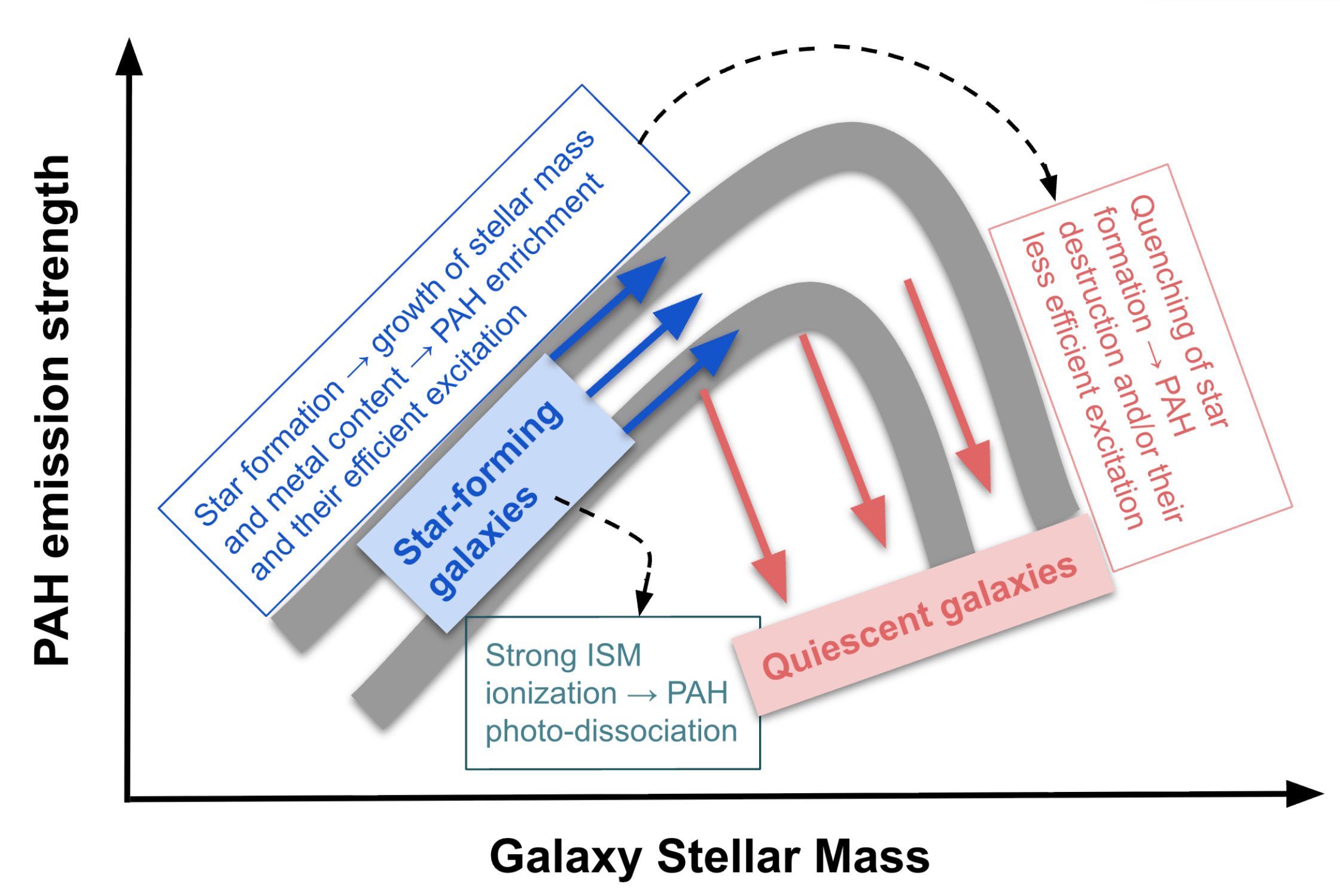}
    \caption{A schematic diagram to demonstrate PAH emission in star-forming and quiescent galaxies. SFGs grow their PAH mass fraction with increasing stellar mass and metallicity, and also produce enough FUV photons to excite the PAH molecules efficiently. However, SFGs with extreme ionizing conditions, which are mostly less massive, show weaker PAH emission as the molecules get photo-dissociated when exposed to such a radiation field. In quenched galaxies, the PAH emission strength drops significantly, as the mechanisms that quenched star formation could also destroy PAH molecules. The inactive environment of QGs, which is inefficient for PAH excitation, further contributes to their weaker PAH emission.}
    \label{fig:fpah_schematic}
\end{figure}

\subsection{PAH emission: from star-forming to quiescent galaxies}
\label{sub_sec:schematic}
The behaviour of f$_{\rm PAH}$ with different galaxy parameters in both SFGs and QGs, as discussed in earlier sections, provides important insights into the evolution of PAH mass fraction as a galaxy evolves with cosmic time. Our results suggest that a galaxy, while being in the star-forming sequence, would gradually enrich PAH as it grows its stellar mass and metal content with cosmic time (Figure \ref{fig:fpah_schematic}). Stars formed in each star-forming episode would contribute to building the total PAH mass in galaxies, while the FUV photons primarily produced by stars from the most recent or ongoing burst would excite the PAH molecules to power the observed emission features. As a result, massive UV-luminous galaxies turn out to be the strongest PAH emitters among the star-forming populations. However, SFGs that host extreme ionizing conditions in the ISM can destroy a fraction of the PAH through photo-dissociation, resulting in weaker PAH emission. Our study infers that the star-forming dwarf galaxies would have a lower PAH mass fraction, which could be driven by both stronger ionizing conditions and lower stellar mass. As demonstrated in Figure \ref{fig:fpah_schematic} and discussed in earlier sections, the quiescent galaxies mostly show relatively lower PAH mass fractions, which could be primarily driven by the efficient destruction of smaller dust particles through the mechanisms that quenched star formation in these systems. These mechanisms may involve a range of physical processes, including AGN-driven radiation and outflows, supernova-driven shocks, morphological quenching, and galaxy mergers, which we plan to investigate in detail in a follow-up study. The inactive star-forming phase of the quiescent galaxies would further result in inefficient excitation of the PAH particles, which in turn would produce weaker PAH emission features during this phase of the evolution. Overall, our study reports a contrasting picture of PAH emission strength in star-forming and quiescent galaxies, providing more scope to explore the scenario in the future.

\section{Summary}
\label{sec:summary}

In this study, we analyze 224 galaxies at $z\sim 0.5-0.8$ in the GOODS-South field to understand PAH mass fraction with varying galaxy stellar mass, FUV luminosity, ISM ionization, and gas-phase metallicity. We define a parameter f$_{\rm PAH}$ as the ratio of the JWST F1280W flux to the mean flux of the F770W and F2550W bands. Using model spectra, we show that f$_{\rm PAH}$ provides a reliable constraint on PAH mass fraction \qpah by probing the strength of PAH 7.7$\mu$m emission in galaxies within our targeted redshift window. We perform SED analysis of the sample galaxies and find a tight correlation between the model \qpah and the observed f$_{\rm PAH}$ parameter, further demonstrating the reliability of f$_{\rm PAH}$ in constraining PAH mass fraction.

The main results of this study are summarized here:

\begin{itemize}

    \item For star-forming galaxies (i.e., 78\% of the total samples), we find a positive correlation between f$_{\rm PAH}$ and observed UV continuum slope ($\beta$), which plausibly signifies that an increase in total dust extinction in galaxies would also mean an increase in PAH mass fraction and vice versa. However, the quiescent galaxies (i.e., 22\% of the total sample) show mixed behavior. The QGs with relatively bluer UV continuum ($\beta<0$) show a similar trend as seen in SFGs, whereas redder QGs ($\beta>0$), most of which are also massive and would host older stellar populations, show low f$_{\rm PAH}$ values.

    \item We notice a strong positive correlation between f$_{\rm PAH}$ and stellar mass in star-forming galaxies, suggesting that the PAH mass fraction is primarily regulated by galaxy stellar mass. Quiescent galaxies, the majority of which are massive (i.e., log(M/[M$_{\odot}$) $>10$), show a lower PAH mass fraction relative to that of SFGs of similar mass. The weaker f$_{\rm PAH}$ value in massive QGs could indicate substantial destruction or removal of smaller dust grains during their evolution. For PAH-poor QGs with no rest-frame UV detection, it could also highlight the importance of energetic FUV photons (which drop significantly in QGs) in heating the PAH dust particles.

    \item For SFGs, f$_{\rm PAH}$ also correlates well with the rest-frame FUV luminosity, indicating the crucial role of FUV photons in exciting PAH molecules in galaxies. We also notice significant scatter in f$_{\rm PAH}$ at a fixed FUV luminosity. The scatter is most likely driven by the galaxy stellar mass, as massive galaxies with similar FUV luminosity show higher f$_{\rm PAH}$. Our results suggest a higher PAH mass fraction in progressively more massive UV-luminous galaxies. We also find that most UVIT-detected QGs exhibit relatively low f$_{\rm PAH}$ despite their moderate UV luminosities and higher stellar masses, suggesting the destruction of PAHs as the primary cause for their low abundance in QGs.

    \item The strength of ISM ionization, measured using the oxygen emission-line ratio O32, plays a crucial role in controlling PAH survival. We notice a strong anti-correlation between f$_{\rm PAH}$ and O32. This observation suggests that PAH molecules exposed to a harsh ionizing ISM radiation field can be destroyed in a large fraction. The fact that the stellar mass of the galaxies also shows an anti-correlation with O32 further suggests that both galaxy mass and ISM ionization strength are important factors in deciding PAH mass fraction in galaxies. However, we also find that, at fixed stellar masses, especially in the low-mass regime, higher O32 values are mostly associated with lower $f_{\rm PAH}$, confirming enhanced PAH photo-dissociation under harsher ionizing conditions.

    \item We find a clear positive correlation between PAH mass fraction and gas-phase metallicity in star-forming galaxies at $z\sim0.65$. Our study further indicates the presence of a threshold metallicity of 12 + log(O/H) $\sim$ 8.45 (Z$\sim$0.7Z$_\odot$), below which the PAH emission strength starts to drop significantly, consistent with findings from both local and high-redshift SFGs. The close agreement of this threshold with previous studies across a wide redshift range suggests that the metallicity required for substantial PAH formation has remained largely unchanged over cosmic time.
    
\end{itemize}

\newpage

\section*{Data Availability}
The JWST data, including JADES \citep{rieke2023} and SMILES \citep{alberts2024,rieke2024} data, were obtained from MAST: 
\citet{10.17909/8tdj-8n28} and \citet{10.17909/et3f-zd57}, respectively. 
The AstroSat/UVIT \citep{saha2024} data were taken from the machine-readable catalogs (Table 5) published online.
The HST data, including HDUV \citep{oesch2018} and 3D-HST \citep{momcheva2016,Brammer2012} data,
were obtained from MAST: \citet{10.17909/t90t2n} and \citet{10.17909/t9jw9z}, respectively.
The Herschel data \citep{elbaz2011} were obtained from Vizier \citep{10.26093/cds/vizier.35330119}. 
Properties for our 224 sample galaxies (Table C.1) are available in a machine-readable format in the online Journal.

\begin{acknowledgments}
This work is based (in part) on observations made with the NASA/ESA/CSA James Webb Space Telescope. The data were obtained from the Mikulski Archive for Space Telescopes at the Space Telescope Science Institute, which is operated by the Association of Universities for Research in Astronomy, Inc., under NASA contract NAS 5-03127 for JWST. This research is also based (in part) on observations made with the AstroSat/Ultra-violet Imaging Telescope, NASA/ESA Hubble Space Telescope, and ESA/Herschel space observatory. C.-C.C. acknowledges support from the National Science and Technology Council of Taiwan (114-2628-M-001-006-MY4), as well as Academia Sinica through the Career Development Award (AS-CDA-112-M02). H.S.B.A gratefully acknowledges support from Academia Sinica through grant AS-PD-1141-M01-2. Finally, we thank the referee for valuable suggestions.

\end{acknowledgments}



\facilities{AstroSat (UVIT), JWST (MIRI, NIRCam), HST (ACS, WFC3), Herschel (PACS), MUSE}

\software{SAOImageDS9 \citep{joye2003}, Matplotlib \citep{matplotlib2007}, Astropy \citep{astropy2013,astropy2018,astropy2022}, CIGALE \citep{boquien2019}}


\appendix

\counterwithin{table}{section}
\counterwithin{figure}{section}

\section{Variation of the MIR emission with dust parameters $\alpha$ and $\gamma$}
\label{app_sed}

\begin{figure*}[h]
    \centering
    \includegraphics[width=5.5in]{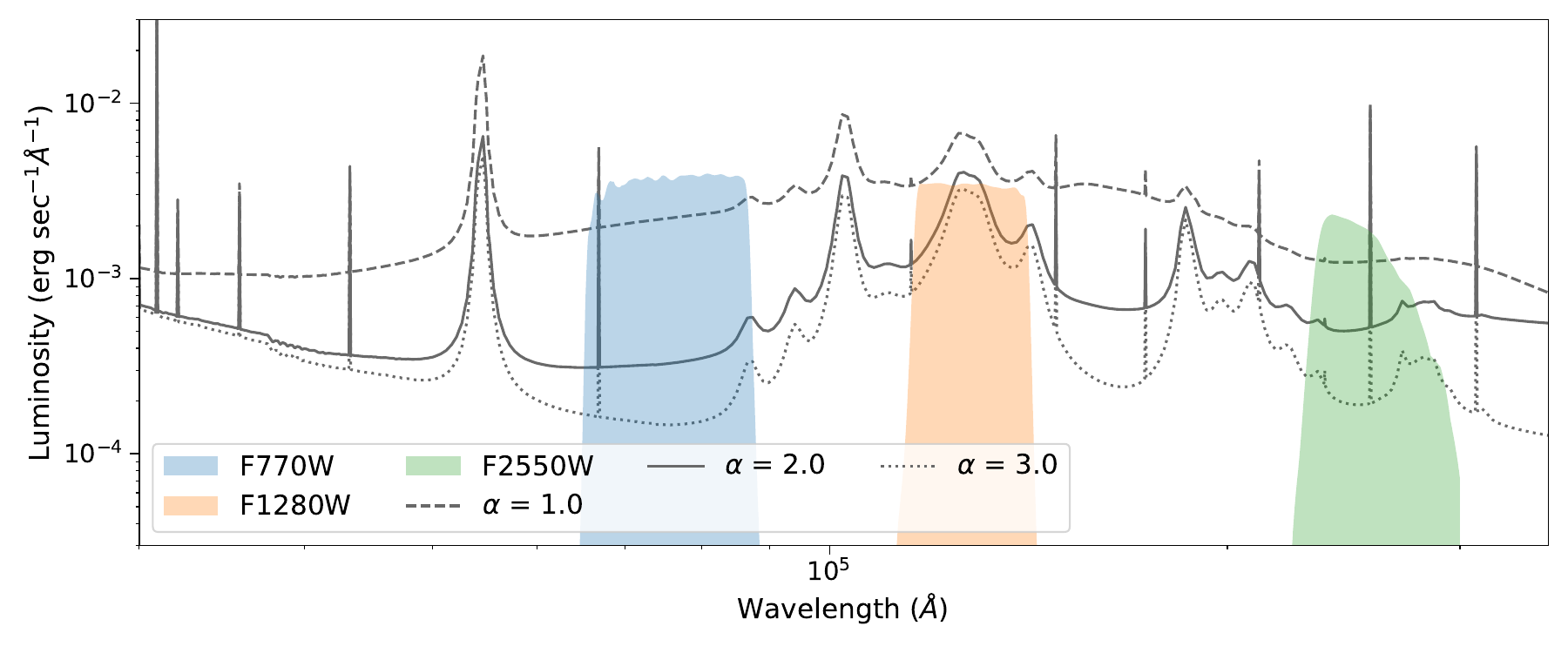}\\
    \includegraphics[width=5.5in]{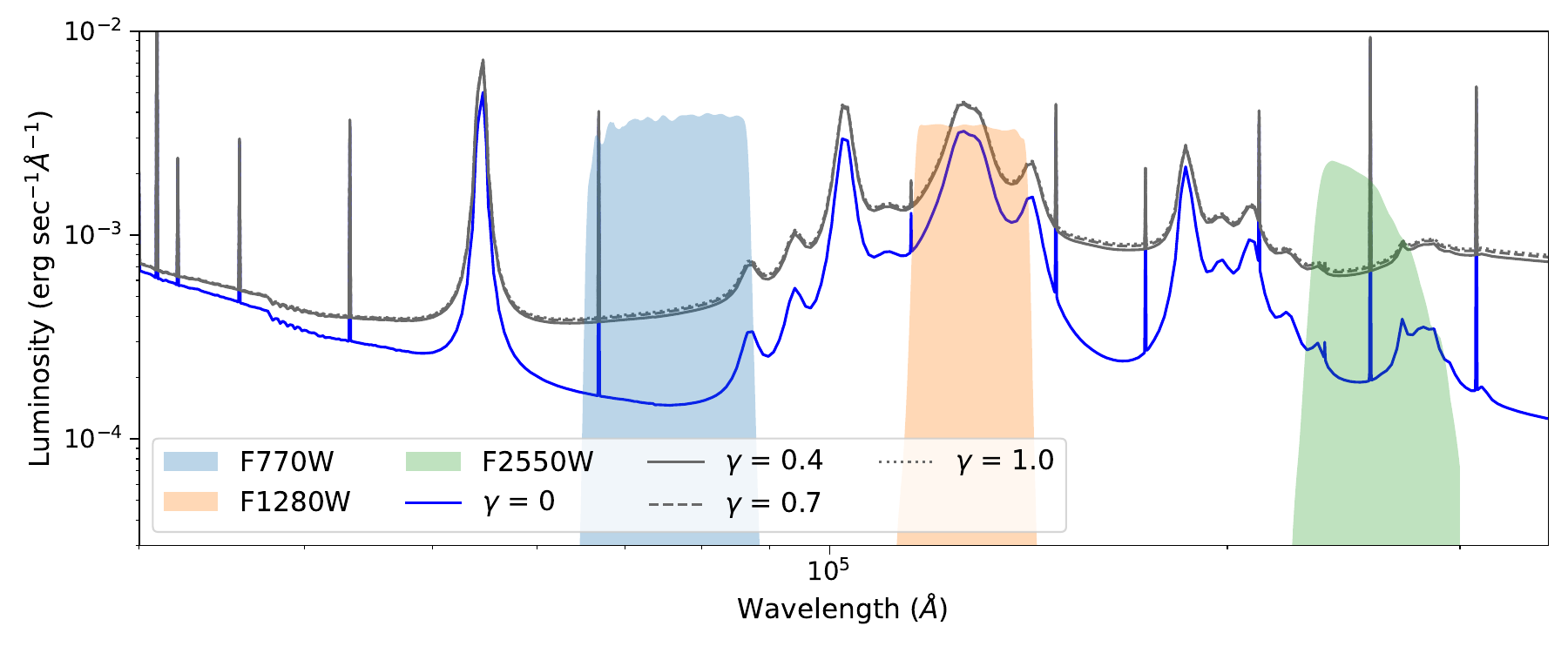}
    \caption{Variation of MIR-continuum and the PAH emission bands with dust parameters $\alpha$ (top) and $\gamma$ (bottom) for an adopted \qpah value of 4.58. Three JWST/MIRI bands that are used to define the parameter f$_{\rm PAH}$ (equation \ref{eq:fpah}) are shown to demonstrate how the variation of $\alpha$ and $\gamma$ can affect f$_{\rm PAH}$.}
    \label{fig:alpha_gamma}
\end{figure*}

\section{AGN populations}
\label{sec:app_agn}
\begin{figure*}[h]
    \centering
    \includegraphics[width=5.5in]{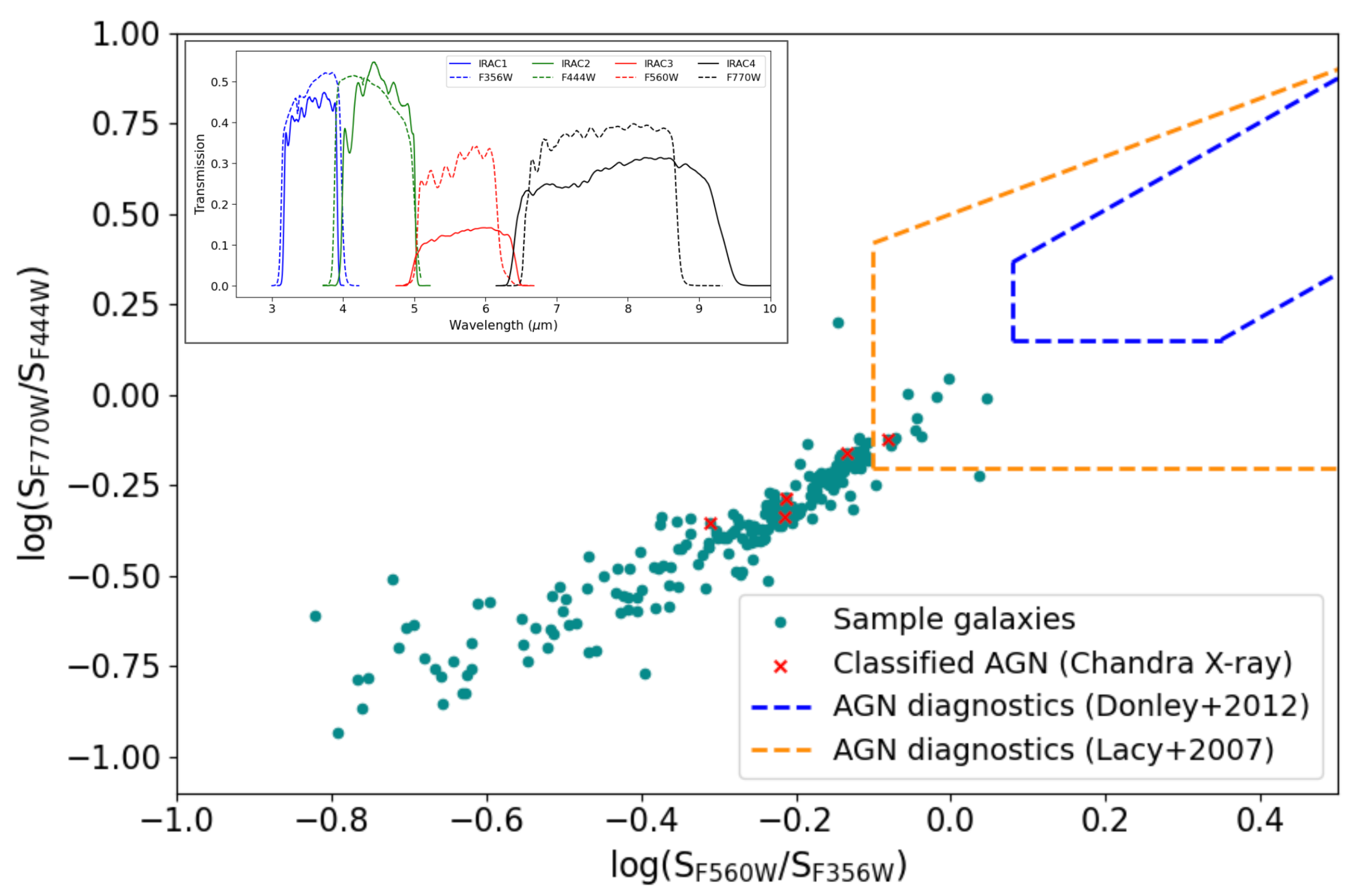}
    \caption{MIR color-color diagnostics to identify potential AGN populations in our sample galaxies. The JWST flux ratios S$_{\rm F770W}$/S$_{\rm F444W}$ and S$_{\rm F560W}$/S$_{\rm F356W}$ respectively probe the S$_{8.0 \mu m}$/S$_{4.5\mu m}$ and S$_{5.8 \mu m}$/S$_{3.6\mu m}$ IRAC flux ratios used in \citet{donley2012} and \citet{lacy2007} to classify AGN populations based on the conditions as indicated using the blue and orange dashed lines, respectively. We find no sample galaxies to fall within the region classified as AGN in the revised MIR color-color diagram by \citet{donley2012}, whereas only 11 sample galaxies qualify as potential AGNs as per \citet{lacy2007}. 5 sample galaxies that are classified as AGN in the Chandra X-ray catalog are also marked. In the inset, we show the filter transmission profiles of four JWST bands, each covering almost the same wavelength range as one of the four IRAC bands used to define the color-color criteria in \citet{donley2012} and \citet{lacy2007}.}
    \label{fig:agn}
\end{figure*}

In order to know the fraction of potential AGN in our selected samples, we first utilize the 7 MS Chandra X-ray source catalog of the GOODS-South field \citep{luo2017}. Following a cross-match, we find only 5 sources classified as AGN among our 224 sample galaxies. We further construct an MIR color-color diagram to identify potential AGN candidates among our sample. \citet{lacy2007} used four Spitzer/IRAC bands to separate AGN populations based on two flux ratios: IRAC3 (5.8$\mu$m)/ IRAC1 (3.6$\mu$m) and IRAC4 (8.0$\mu$m)/ IRAC2 (4.5$\mu$m). The conditions provided in \citet{lacy2007} were further updated by \citet{donley2012} to separate AGNs from star-forming galaxies. In Figure \ref{fig:agn} (inset), we show the transmission curves of four IRAC bands along with four bands from JWST (two NIRCam and 2 MIRI filters). Each of the JWST bands, for which we have flux measurements, covers almost the same wavelength range as one corresponding IRAC band that was used by \citet{donley2012} and \citet{lacy2007} to identify AGN populations. We estimate two flux ratios S$_{\rm F770W}$/S$_{\rm F444W}$ and S$_{\rm F560W}$/S$_{\rm F356W}$, which closely represent S$_{\rm IRAC4}$/S$_{\rm IRAC2}$ and S$_{\rm IRAC3}$/S$_{\rm IRAC1}$, respectively, for all the sample galaxies and shown in Figure \ref{fig:agn}. We find no galaxy to satisfy the IRAC-based MIR color-color criteria to qualify as an AGN as specified in \citet{donley2012}, while the number of potential AGNs becomes only 11 when we consider the diagnostics from \citet{lacy2007}. We note that the JWST-based MIR color-color conditions may differ slightly from those quoted in both \citet{donley2012} and \citet{lacy2007}. However, it would not make a significant change to the identified number of possible AGNs. Therefore, both the X-ray cross-match and the MIR color-color classification imply a negligible fraction ($<7\%$) of potential AGN populations in our samples.

\section{Catalog of sample galaxies}
\label{app_catalog}
We provide a part of the source catalog with their position, redshift, f$_{\rm PAH}$, \qpah, stellar mass, and sSFR values in Table \ref{tab:catalog}. The full version of the table would be made available online along with the paper.
\begin{table*}
\begin{flushleft}
\caption{Properties of the 224 sample galaxies$^*$}
\label{tab:catalog}
\begin{tabular}{p{1cm}p{1.5cm}p{1.5cm}p{1.5cm}p{1cm}p{1.cm}p{1cm}p{1cm}p{1.2cm}p{1.2cm}p{1.2cm}p{1.2cm}}
\hline

ID & RA & Dec & $z$ & f$_{\rm PAH}$ & f$_{\rm PAH}$ error & \qpah & \qpah error & Mass & Mass error & log(sSFR) & sSFR error \\

(1) & (2) & (3) & (4) & (5) & (6) & (7) & (8) & (9) & (10) & (11) & (12) \\\hline

189671 & 53.159816 & -27.841338 & 0.6010 & 4.63 & 0.29 & 5.43 & 1.24 & 6.4E+09 & 1.5E+09 & -9.01 & 8.7E-10\\
173928 & 53.150880 & -27.868323 & 0.7607 & 2.58 & 0.10 & 4.07 & 0.70 & 8.4E+10 & 9.9E+09 & -11.27 & 3.9E-12\\
202891 & 53.168404 & -27.797871 & 0.5403 & 3.84 & 2.09 & 4.67 & 1.36 & 1.6E+09 & 2.8E+08 & -10.41 & 3.8E-11\\
190562 & 53.093169 & -27.838422 & 0.7228 & 4.18 & 0.67 & 5.06 & 1.13 & 8.5E+09 & 2.0E+09 & -10.38 & 7.4E-11\\
190219 & 53.086248 & -27.839506 & 0.7297 & 1.74 & 0.90 & 3.36 & 1.82 & 2.0E+09 & 4.3E+08 & -8.83 & 7.4E-10\\
176497 & 53.121560 & -27.864772 & 0.5060 & 1.62 & 0.11 & 1.12 & 0.03 & 1.0E+09 & 3.4E+08 & -8.70 & 8.7E-10\\
207811 & 53.196994 & -27.780593 & 0.7419 & 2.05 & 0.74 & 3.71 & 2.16 & 1.0E+09 & 2.0E+08 & -8.97 & 6.9E-10\\
190738 & 53.113983 & -27.837759 & 0.7710 & 4.34 & 0.23 & 5.21 & 0.88 & 1.7E+10 & 3.5E+09 & -9.08 & 8.2E-10\\
199958 & 53.137538 & -27.808167 & 0.6640 & 1.65 & 0.23 & 5.18 & 1.64 & 2.2E+09 & 3.7E+08 & -9.11 & 6.8E-10\\
200196 & 53.118651 & -27.807373 & 0.6690 & 5.63 & 0.13 & 2.27 & 0.57 & 1.0E+10 & 1.7E+09 & -8.76 & 6.3E-10\\
208912 & 53.151041 & -27.777304 & 0.7263 & 1.40 & 0.26 & 4.85 & 1.80 & 8.3E+08 & 1.4E+08 & -9.25 & 4.9E-10\\
109276 & 53.190751 & -27.803576 & 0.5336 & 2.35 & 0.16 & 4.44 & 0.97 & 2.0E+10 & 3.1E+09 & -11.16 & 8.5E-12\\
204639 & 53.193680 & -27.790891 & 0.5250 & 0.81 & 0.43 & 3.48 & 2.24 & 6.7E+08 & 1.3E+08 & -9.54 & 3.7E-10\\
218780 & 53.178614 & -27.739219 & 0.5250 & 3.52 & 2.33 & 5.84 & 1.14 & 8.5E+08 & 1.8E+08 & -9.11 & 6.4E-10\\
183156 & 53.137200 & -27.855687 & 0.6644 & 1.09 & 0.29 & 0.99 & 0.92 & 1.2E+09 & 3.4E+08 & -8.79 & 9.0E-10\\
196915 & 53.161704 & -27.819254 & 0.6665 & 5.21 & 0.36 & 5.44 & 1.58 & 7.3E+09 & 1.5E+09 & -9.89 & 1.9E-10\\
99148 & 53.143897 & -27.817991 & 0.7031 & 0.31 & 0.41 & 0.96 & 0.77 & 6.4E+08 & 1.4E+08 & -8.78 & 8.0E-10\\
204984 & 53.168010 & -27.789666 & 0.6190 & 3.81 & 0.42 & 5.45 & 1.57 & 1.8E+09 & 4.0E+08 & -8.98 & 8.2E-10\\
200847 & 53.151188 & -27.805563 & 0.6652 & 1.14 & 1.16 & 3.67 & 2.04 & 1.1E+09 & 2.7E+08 & -9.16 & 5.8E-10\\
204264 & 53.169464 & -27.791938 & 0.6670 & 0.50 & 0.03 & 0.47 & 0.00 & 4.2E+10 & 2.1E+09 & -11.89 & 3.9E-13\\
203023 & 53.173618 & -27.797356 & 0.7006 & 0.96 & 0.20 & 4.26 & 1.87 & 2.3E+09 & 3.5E+08 & -9.33 & 4.5E-10\\
186758 & 53.135727 & -27.848307 & 0.6628 & 5.32 & 0.49 & 4.26 & 1.10 & 1.5E+10 & 3.4E+09 & -8.95 & 6.8E-10\\
225651 & 53.157792 & -27.844292 & 0.5196 & 0.68 & 0.41 & 0.52 & 0.22 & 1.8E+08 & 3.7E+07 & -8.85 & 3.5E-10\\
198400 & 53.110355 & -27.814327 & 0.7612 & 0.92 & 0.32 & 3.62 & 2.15 & 2.4E+09 & 4.7E+08 & -9.53 & 4.2E-10\\
176480 & 53.088585 & -27.864132 & 0.6719 & 1.96 & 0.65 & 5.01 & 1.74 & 1.1E+09 & 2.6E+08 & -8.72 & 6.4E-10\\
219504 & 53.161186 & -27.734074 & 0.6845 & 0.92 & 0.45 & 3.84 & 2.16 & 1.4E+09 & 2.7E+08 & -10.06 & 1.8E-10\\
133285 & 53.129334 & -27.770893 & 0.7091 & 1.00 & 0.63 & 1.77 & 1.56 & 1.5E+09 & 2.7E+08 & -8.85 & 7.1E-10\\
184784 & 53.122548 & -27.852727 & 0.7350 & 3.59 & 0.09 & 2.68 & 0.46 & 5.0E+10 & 1.1E+10 & -10.33 & 1.4E-10\\
195504 & 53.186264 & -27.822994 & 0.5220 & 3.66 & 0.33 & 5.58 & 1.03 & 1.3E+10 & 2.6E+09 & -9.40 & 4.4E-10\\
194297 & 53.095055 & -27.826199 & 0.7272 & 2.82 & 0.63 & 3.39 & 0.98 & 3.9E+09 & 1.0E+09 & -8.76 & 7.9E-10\\
...\\\hline

\end{tabular}
\tablenotetext{*}{(The full table is available in machine-readable form)}

\textbf{Note.} Table columns: (1) NIRCam source ID as mentioned in JADES catalog \citep{rieke2023}, (2) Right Ascension (RA) in degrees, (3) Declination in degrees, (4) Redshift ($z_{\rm best}$) from the 3D-HST catalog \citep{momcheva2016}, (5) observed f$_{\rm PAH}$ value, (6) error in f$_{\rm PAH}$, (7) SED-derived \qpah value, (8) error in \qpah, (9) SED-derived stellar mass in M$_{\odot}$, (10) error in stellar mass in M$_{\odot}$, (11) SED-derived specific star formation rate (sSFR) in log(sSFR) [yr$^{-1}$], (12) error in sSFR [yr$^{-1}$].
\end{flushleft}
\end{table*}

\end{document}